\documentclass[11pt, letterpaper]{article}

\usepackage{times}
\usepackage{wrapfig}
\usepackage{sectsty}
\sectionfont{\large}
\subsectionfont{\normalsize}
\subsubsectionfont{\small}
\usepackage{listings}
\usepackage{caption}
\usepackage{subcaption}

\usepackage{amsfonts}
\usepackage{microtype}
\usepackage{graphicx}

\usepackage{booktabs}

\usepackage[most]{tcolorbox}
\usepackage{fontawesome5} %

\usepackage{xcolor}

\definecolor{system}{HTML}{2CA02C} %
\definecolor{user}{HTML}{9467BD}   %
\definecolor{agent}{HTML}{FF7F0E}  %

\definecolor{light}{HTML}{D3D3D3} 
\tcbset{
  bubble/.style 2 args={
    enhanced jigsaw,         %
    breakable,               %
    segmentation style={solid}, %
    pad at break*=1mm,       %
    sharp corners=south,
    rounded corners,
    arc=2mm,
    colback=#1!8,
    colframe=#1!60!black,
    boxrule=0.5pt,
    left=3mm, right=3mm, top=2mm, bottom=2mm,
    fontupper=\small,
    fonttitle=\small,
    before skip=6pt, after skip=12pt,
    width=\linewidth,
    title={#2}
  }
}

\tcbset{
  systemblock/.style={bubble={system}{\faCog\; system:}},
  userblock/.style={bubble={user}{\faUser\; user:}},
  agentblock/.style={bubble={agent}{\faRobot\; agent:}},
}

\newenvironment{system}[1][]{%
  \par\addvspace{\medskipamount}%
  \begin{tcolorbox}[
    systemblock,
    title={\faCog\; \textit{system}:\if\relax\detokenize{#1}\relax\else\ \ #1\fi},
    title after break={\faCog\; \textit{system} (cont.):\if\relax\detokenize{#1}\relax\else\ \ #1\fi}
  ]%
}{%
  \end{tcolorbox}\par\addvspace{\medskipamount}%
}

\newenvironment{user}[1][]{%
  \par\addvspace{\medskipamount}%
  \begin{tcolorbox}[
    userblock,
    title={\faUser\; \textit{user}:\if\relax\detokenize{#1}\relax\else\ \ #1\fi},
    title after break={\faUser\; \textit{user} (cont.):\if\relax\detokenize{#1}\relax\else\ \ #1\fi}
  ]%
}{%
  \end{tcolorbox}\par\addvspace{\medskipamount}%
}

\newenvironment{agent}[1][]{%
  \par\addvspace{\medskipamount}%
  \begin{tcolorbox}[
    agentblock,
    title={\faRobot\; \textit{agent}:\if\relax\detokenize{#1}\relax\else\ \ #1\fi},
    title after break={\faRobot\; \textit{agent} (cont.):\if\relax\detokenize{#1}\relax\else\ \ #1\fi}
  ]%
}{%
  \end{tcolorbox}\par\addvspace{\medskipamount}%
}

\usepackage{multicol}
\usepackage{algorithm}
\usepackage{algpseudocode}
\usepackage{enumitem}

\usepackage{setspace}
\usepackage{natbib}

\usepackage{amsmath}
\usepackage{amssymb}
\usepackage{mathtools}
\usepackage{amsthm}

\usepackage[colorlinks,
            linkcolor=red,
            anchorcolor=blue,
            citecolor=blue
            ]{hyperref}
\usepackage[capitalize,noabbrev]{cleveref}

\DeclareFontFamily{U}{mathb}{}
\DeclareFontShape{U}{mathb}{m}{n}{
      <-6> mathb5
      <6-7> mathb6
      <7-8> mathb7
      <8-9> mathb8
      <9-10> mathb9
      <10-12> mathb10
      <12-> mathb12
}{}
\DeclareSymbolFont{mathb}{U}{mathb}{m}{n}

\DeclareMathSymbol{\drsh}{3}{mathb}{"EB}

\theoremstyle{plain}
\newtheorem{theorem}{Theorem}[section]

\newtheorem{definition}[theorem]{Definition}

\theoremstyle{definition}

\newtheorem{remark}[theorem]{Remark}

\usepackage{smile}
\usepackage{makecell}

\usepackage{amsmath,amsfonts,bm}

\def\1{\bm{1}}

\def\vf{{\bm{f}}}

\def\vh{{\bm{h}}}

\def\vp{{\bm{p}}}

\def\evf{{f}}

\def\evp{{p}}

\def\evu{{u}}
\def\evv{{v}}

\def\evx{{x}}

\DeclareMathAlphabet{\mathsfit}{\encodingdefault}{\sfdefault}{m}{sl}
\SetMathAlphabet{\mathsfit}{bold}{\encodingdefault}{\sfdefault}{bx}{n}

\def\gA{{\mathcal{A}}}

\def\gF{{\mathcal{F}}}
\def\gG{{\mathcal{G}}}

\def\gK{{\mathcal{K}}}

\def\gM{{\mathcal{M}}}
\def\gN{{\mathcal{N}}}
\def\gO{{\mathcal{O}}}

\def\gR{{\mathcal{R}}}

\def\gV{{\mathcal{V}}}
\def\gW{{\mathcal{W}}}

\def\sR{{\mathbb{R}}}

\def\emLambda{{\Lambda}}

\DeclareMathOperator*{\argmin}{arg\,min}

\newcommand{\interior}[1]{%
  {\kern0pt#1}^{\mathrm{o}}%
}

\newcommand\extrafootertext[1]{%
    \bgroup
    \renewcommand\thefootnote{\fnsymbol{footnote}}%
    \renewcommand\thempfootnote{\fnsymbol{mpfootnote}}%
    \footnotetext[0]{#1}%
    \egroup
}
\usepackage[textsize=tiny]{todonotes}
\usepackage{fullpage}

\usepackage{xspace}
\newcommand{\result}{\textless result\textgreater\xspace}
\newcommand{\closeresult}{\textless/result\textgreater\xspace}

\usepackage{float}
\usepackage{titlesec}

\titleformat{\section}
  {\normalfont\large\bfseries}{\thesection}{1em}{}
\titleformat{\subsection}
  {\normalfont\normalsize\bfseries}{\thesubsection}{1em}{}
\titleformat{\subsubsection}
  {\normalfont\normalsize\bfseries}{\thesubsubsection}{1em}{}
\titlespacing*{\section}{0pt}{5pt plus 1pt minus 1pt}{2pt plus 1pt minus 1pt}

\titlespacing*{\subsection}{0pt}{4pt plus 1pt minus 1pt}{2pt plus 1pt minus 1pt}

\titlespacing*{\subsubsection}{0pt}{4pt plus 1pt minus 1pt}{2pt plus 1pt minus 1pt}

\titlespacing*{\paragraph}{0pt}{2pt plus 1pt minus 1pt}{4pt plus 1pt minus 1pt}

\labelwidth\leftmargini\advance\labelwidth-\labelsep \labelsep 5pt

\def\@listi{\leftmargin\leftmargini}
\def\@listii{\leftmargin\leftmarginii
   \labelwidth\leftmarginii\advance\labelwidth-\labelsep
   \topsep 2pt plus 1pt minus 0.5pt
   \parsep 1pt plus 0.5pt minus 0.5pt
   \itemsep \parsep}
\def\@listiii{\leftmargin\leftmarginiii
    \labelwidth\leftmarginiii\advance\labelwidth-\labelsep
    \topsep 1pt plus 0.5pt minus 0.5pt
    \parsep \z@ \partopsep 0.5pt plus 0pt minus 0.5pt
    \itemsep \topsep}
\def\@listiv{\leftmargin\leftmarginiv
     \labelwidth\leftmarginiv\advance\labelwidth-\labelsep}
\def\@listv{\leftmargin\leftmarginv
     \labelwidth\leftmarginv\advance\labelwidth-\labelsep}
\def\@listvi{\leftmargin\leftmarginvi
     \labelwidth\leftmarginvi\advance\labelwidth-\labelsep}
     
\usepackage{authblk}

\title{\textbf{\large{LLM-Guided Reinforcement Learning with Representative Agents for Traffic Modeling}}}

\author
{\normalsize
Hanlin Sun$^{1}$
\qquad Jiayang Li$^{1,\star}$
}
\affil{$^1$\textit{\small Department of Data and Systems Engineering, The University of Hong Kong}.}

\date{}

\begin{document}

\extrafootertext{$^\star$Corresponding author; email: \texttt{jiayangl@hku.hk}.
}

\maketitle

\begin{abstract}

Large language models (LLMs) are increasingly used as behavioral proxies for self-interested travelers in agent-based traffic models. Although more flexible and generalizable than conventional models, the practical use of these approaches remains limited by scalability due to the cost of calling one LLM for every traveler. Moreover, it has been found that LLM agents often make opaque choices and produce unstable day-to-day dynamics. To address these challenges, we propose to model each homogeneous traveler group facing the same decision context with a single representative LLM agent who behaves like the population's average, maintaining and updating a mixed strategy that coincides with the group's aggregate flow proportions. Each day, the LLM reviews the travel experience and flags commuting options (e.g., routes) with positive reinforcement that they hope to use more often, and an interpretable update rule then converts this judgment into strategy adjustments using a tunable (progressively decaying) step size. The representative-agent design improves scalability, while the separation of reasoning from updating clarifies the decision logic while stabilizing learning. In classic traffic assignment settings, where UE provides the natural benchmark, the proposed dynamics remain stable and approach near-UE states. In richer settings, the generated dynamics remain stable,  producing sensible and interpretable behavioral regularities, for example, the decoy-like context effects in toll versus non-toll road selection, a reasonable balance between travel time and denied-boarding risk in oversaturated transit systems, and higher willingness-to-pay for convenience among higher-income travelers when choosing between driving, transit, and park-and-ride options.

\vspace{5pt}
    \vspace{1em}
\end{abstract}

\vspace{1em}

\section{Introduction}

Growing evidence has demonstrated that current large language models (LLMs) can already imitate many aspects of human intelligence \citep{wei2022chain, meta2022human, park2023generative}. This progress has inspired a range of new research questions in game theory, including whether LLM agents can act as rational decision-makers in strategic interactions. In this direction, many have shown that non-cooperative games played by LLM agents can evolve in ways largely resembling the behavior of rational human players \citep[e.g.,][]{guo2023gpt, park2023generative, han2023guinea, williams2023epidemic, akata2025playing}. Building on these insights, the proposed research seeks to explore the potential of LLMs in modeling the behavior of self-interested travelers to advance a novel LLM-based approach to traffic modeling.

\subsection{Conventional traffic modeling}

For decades, conventional models have provided powerful analytical tools for traffic planning. Most conventional models --- whether equilibrium-based, describing travelers' stable behavior, or day-to-day, capturing how their behaviors evolve over time --- rely on explicitly defined behavioral rules for route selection. User equilibrium (UE) and stochastic UE (SUE), for example, respectively rest on the assumptions that travelers will not change their routes once no better route exists \citep{wardrop1952road} and no better alternative is ``believed" to exist, subject to random perception errors \citep{daganzo1977stochastic}. Among day-to-day models, a major class relies on a ``revision protocol" that specifies the probability of a traveler switching from their current route to another, based on costs experienced in the previous day \citep{smith1984stability, sandholm2010population}.
Another line of models assumes travelers' route choice is determined by their evaluation of all available routes, which may be either a weighted average \citep{horowitz1984stability}  or an accumulation \citep{li2024wardrop, li2024day} of historical costs. 

{However, these models may become less adequate when generalized to richer behavioral settings where the travel experience contains multiple attributes, contextual descriptions, and traveler-profile information.}
\begin{itemize}[leftmargin=*, topsep=0pt, itemsep=2pt, parsep=0pt]
    \item First, conventional models often require multi-attribute travel experiences to be converted into a scalar fixed-form generalized cost, which may oversimplify how disutility is evaluated. Take transit trips, for instance, where a common approach is to represent each traveler's generalized cost as a weighted sum of total time and money. Yet, this approach overlooks how a single minute can be experienced very differently --- whether it is spent walking to the station, waiting at the station, or riding (with or without a seat) --- and how the perceived disutility of time does not accumulate linearly --- for instance, anxiety may intensify once waiting time exceeds a certain threshold. Monetary perception can be equally complex, for example, the ``zero-price effect" \citep{ShampanierMazarAriely2007}, where the jump from \$0 to any positive price triggers a disproportionately large drop in utility compared to equal increments among positive prices. 
    \item Second, conventional models often simplify option comparison by assuming that travelers minimize a pre-specified scalar cost, which may oversimplify how they compare alternatives with multiple attributes. The decoy effect is one example: when an asymmetrically dominated alternative is introduced, the option that dominates it may become more attractive, even if the new alternative is rarely chosen \citep{huber1982adding}. Other comparison effects can also arise. For example, a compromise option often becomes attractive because it lies between two extremes \citep{simonson1989choice}. There is no single standard scalar cost-minimization structure that automatically captures all these comparison patterns without the modeler specifying the relevant behavioral mechanism in advance. Further, the difficulty is that the modeler may not know ex ante which comparison structure travelers will invoke in a new travel context.

    \item Third, conventional models often represent traveler heterogeneity through a limited set of predefined numerical parameters, which may oversimplify how personal profiles shape perceived utilities and preferences. In reality, individual attributes can fundamentally alter how travel experiences are evaluated. For example, an elderly passenger may avoid a transit option without guaranteed seating, while a student with a tight class schedule may experience heightened anxiety when faced with an unexpectedly long wait at a bus stop. Some of these relationships can be quantified; for instance, higher-income travelers are often assumed to value time more than money, which is commonly modeled through a value-of-time parameter \citep{yang2004multi}. Yet many others, such as perceptions of comfort, convenience, or reliability, are far more difficult to formalize within conventional mathematical frameworks.

\end{itemize}

\subsection{LLM-driven day-to-day models}

Recently, researchers have proposed fully \textit{LLM-driven} day-to-day models in which the entire learning process is delegated to LLMs: each day, travel experiences are conveyed to the agents in plain language, and the agents themselves are asked to directly decide how to revise their route choices \citep{wang2024ai, liu2025toward}. Compared with conventional models, the LLM-driven framework offers several advantages. (1) LLMs can mimic travelers' behavior without relying on limited rules or simplified assumptions of conventional models, since they are pre-trained on vast human-generated data that embeds various decision-making patterns. (2) LLMs can directly interpret rich, \textit{plain-language} descriptions of travel experiences that reflect a broad range of numeric/non-numeric factors without the need to convert every attribute (such as time, money, comfort, and risk) into numerical values and aggregate them into a generalized cost. (3) LLMs can be endowed with detailed traveler profiles, allowing them to represent populations with heterogeneous preferences, socio-demographic characteristics, or behavioral tendencies \citep{liu2025toward}. Despite these promising prospects, fully LLM-driven agent-based models face many challenges.
\begin{itemize}[leftmargin=*, topsep=0pt, itemsep=2pt, parsep=0pt]
    \item  \textbf{Poor scalability}. In agent-based models, to accurately model system-level traffic flows and ensure statistical reliability, it is typically necessary to generate a large population of agents for each origin-destination (OD) pair in the network \citep{macal2010tutorial}. Accordingly, implementing such a simulation at scale shall present significant computational challenges, as on each simulated day, every agent must invoke the LLM at least once to generate a route decision based on the past travel experiences expressed in natural language. Thus, even for a modestly sized network with hundreds of OD pairs, the total number of tokens processed on each day can be extremely high, leading to heavy time costs for inference, as well as considerable memory and hardware requirements for executing model computations \citep{liu2025toward}. 
    \item \textbf{Weak reliability.} In conventional traffic models, every travel decision can be traced to an \textit{explicit} behavioral rule. By contrast, although LLM agents may produce behaviors close to those of humans, their decision-making is harder to audit, and studies show that LLMs often fail to provide logically consistent or strategically coherent explanations for their chosen strategies \citep{lin2024gamebot, herr2024large, maiya2025explaining}, which undermines their suitability as a benchmark for policy evaluation. Stability is another concern: unlike classical day-to-day models, which admit rigorous convergence guarantees to equilibria such as UE or SUE, fully LLM-driven simulations offer no such assurances. In practice, the randomness inherent in generative LLMs can further amplify fluctuations in agent behavior. As a result, the dynamics generated by LLM often fail to settle into a stable state \citep{wang2024ai}, further limiting their reliability.
\end{itemize}

\subsection{Our contribution}

Having discussed the limitations of current approaches, we are led to the following question: \textit{How can the semantic and reasoning capabilities of LLMs be harnessed to build a reliable and efficient framework for modeling complex traffic systems?}

To develop such a framework, we draw on the representative-agent idea widely used in macroeconomics and population-game modeling \citep{hartley2002representative}. In plain terms, a representative agent is a hypothetical decision-maker who stands in for an entire homogeneous population. The rationale is that, when a population shares the same profile, objective, and constraints, a single representative agent's mixed strategy can reasonably describe the proportions of the population adopting different pure strategies. In transportation network analysis, many existing day-to-day learning models similarly assume that, within a homogeneous traveler class, travelers adopt the same mixed strategy \citep{krichene2015convergence,li2024wardrop}, or can be equivalently interpreted in this way \citep{horowitz1984stability}. Building on this population-level modeling tradition, we introduce a representative-LLM construction for day-to-day traffic modeling. Specifically,  we propose to introduce --- for each group of \textit{homogeneous} travelers facing the \textit{same} decision context --- a single LLM agent as their representative. The \emph{representative LLM agent} is initialized with a natural-language description of the decision context and the profile of the class it represents, and is tasked with maintaining and progressively updating a \emph{mixed} strategy over available actions (route or mode) based on day-to-day experience. Each day, the travelers it represents are assumed to act independently by sampling from this mixed strategy. Consequently, the framework moves from one LLM per traveler to one per class to substantially reduce the number of LLM calls processed per simulated day.

To improve the reliability of the dynamics generated by representative LLM agents' learning, we propose an \emph{LLM-guided} mechanism rather than leaving the strategy updates entirely to the LLM. One way to integrate LLMs with conventional learning algorithms is to use them to construct a behavioral representation. For example, one may ask the LLM to design a scalar utility function \citep[e.g.,][]{behari-etal-2024-decision,verma2024_balancing_act}, or ask the LLM to generate synthetic preference data, and then fit a downstream utility model using those generated data \citep[e.g.,][]{alamdari2024jump}. Related work also attempts to combine LLM reasoning with mathematical model discovery: \citet{grayeli2024symbolic} use LLM-derived concepts to guide the discovery of mathematical relationships from data, while \citet{zhao2025personalized} combine learned group-level symbolic utility functions with individual-level semantic adaptation. Our framework addresses a complementary question: how the LLM's direct evaluation of travel experience can be translated into repeated strategy updates without first constructing an intermediate utility model.
Specifically, our mechanism follows the \textit{law of effect} \citep{thorndike1927law}, which states that positively reinforced actions (i.e., those followed by satisfying outcomes) are more likely to be repeated. In behavioral economics, any learning model designed in accordance with the law of effect is typically referred to as a reinforcement learning (RL) model. In the spirit of RL, we propose to decompose an LLM agent's strategy update into two stages. First, the LLM agent receives the travel costs, reflects on the experience, and, using its reasoning ability, identifies the set of routes that yielded positive reinforcement. Second, an explicit, predefined rule increases the probabilities of those routes with \textit{tunable} learning rates. The design thus preserves the core strengths of LLM-based approaches while improving interpretability and stability. On the one hand, the LLM is used where it excels, specifically as a semantic evaluator of rich, multi-attribute travel experiences expressed in natural language and tailored to traveler profiles. On the other hand, the quantitative update follows a transparent rule, and a progressively decaying learning rate promotes convergence.

The proposed framework is first evaluated under a classic traffic assignment setting, where travelers between each OD pair are homogeneous and experience only time-related costs. { This experiment serves as a consistency check. Specifically, if the framework were unstable or deviated substantially from UE (the benchmark) in this basic setting, its use in more complex settings would be less convincing. Across three tested networks, including the Sioux-Falls network, we found that the proposed dynamics remain stable and quickly approach near-UE states. By contrast, the fully LLM-driven baselines, in which the entire strategy update is delegated to the LLM, exhibit persistent instability and do not reliably approach UE. This comparison highlights the importance of the proposed mechanism where the LLM is only used as a semantic evaluator of travel experience.}

We then examine richer scenarios in which travelers must reason over multiple travel attributes rather than a single scalar travel time.  In a highway tolling scenario, travelers choose between toll and non-toll roads, and the introduction of an inferior toll road generates a decoy-like shift toward the original toll road, which illustrates how the framework can represent context-dependent comparison among alternatives. In a transit scenario with potential denied boarding due to vehicle capacity constraints, the framework generates a non-monotonic choice response to changing travel-time outcomes, qualitatively consistent with human experimental findings; comparisons with conventional risk models further illustrate how the LLM-guided agents adapt their choices to the consequences of denied boarding without requiring a pre-specified scalar risk measure. Finally, in a multi-modal commuting scenario, travelers from different income groups choose among transit, driving, and park-and-ride. The resulting patterns are stable and interpretable: lower-income travelers place greater weight on monetary savings; for both higher-income and middle-income travelers, park-and-ride as an intermediate option becomes relatively attractive. This setting also provides several useful sensitivity checks. In particular, it is found that the run-to-run variation of the final mode shares decreases when richer background and profile information is provided to the representative LLM agents, which suggests that well-specified traveler context can improve the stability of LLM-guided predictions.

\subsection{Organization}
The remainder of the paper is organized as follows. Section~\ref{sec:related} reviews the relevant literature. Section~\ref{sec:framework} presents the proposed framework. { Section~\ref{sec:application-tap} applies the framework to the classic traffic assignment problem and Section~\ref{sec:application-rich} demonstrates its use in richer settings.} Finally, Section~\ref{sec:conclusion} concludes and highlights directions for future research.

\section{Related Studies}
\label{sec:related}

We will first review conventional day-to-day traffic assignment models in transportation (Section~\ref{sec:day-to-day}) and reinforcement learning models in behavioral economics (Section~\ref{sec:learning}), before turning to the emerging research on LLM-based game simulation (Section~\ref{sec:llm}).

\subsection{Day-to-day models in transportation}
\label{sec:day-to-day}

Day-to-day (DTD) models capture how travelers iteratively adjust their route choices in response to their experience.  Broadly, they are built on two mechanisms: the first directly switches travelers between routes according to newly realized costs; the second centers on their explicit evaluation of all available routes, which dictates their route choice on each day and is progressively updated.

At the core of the first mechanism is a revision protocol that specifies how travelers switch between routes after observing costs. Representative examples include best-response dynamics, where travelers switch only to the previously lowest-cost route \citep{gilboa1991social}; pairwise comparison dynamics, where travelers may switch from a higher-cost route to a lower-cost one \citep{smith1984stability}; imitative dynamics, where cheaper and more popular routes become more attractive \citep{bjornerstedt1994nash}; and excess-payoff dynamics, where travelers reallocate toward routes whose costs are below the current average \citep{sandholm2005excess,yang2009day}. Under appropriate conditions, these revision-protocol models converge to UE. Extensions incorporating logit choice rules have also been shown to converge to SUE \citep{smith2016route,xiao2019day}. Unlike revision-protocol models, the second mechanism describes how travelers form and update their evaluations of routes based on past experiences. \citet{horowitz1984stability}, for example, proposed that on a given day, a traveler may assign a value to a route using a weighted average. This line of modeling has since been extensively explored in the transportation literature, with subsequent efforts focusing on enriching the behavioral assumptions \citep[e.g.,][]{cascetta1989stochastic, cantarella1995dynamic, watling1999stability, watling2003dynamics}. Because traffic conditions in \citet{horowitz1984stability}'s formulation are subject to perception errors, the resulting DTD dynamics converge to an SUE \citep{daganzo1977stochastic}. More recently, \citet{li2024wardrop} introduced a similar model that converges to UE, which introduces a crucial modification to \citet{horowitz1984stability}'s model: rather than valuing routes by averaging historical costs, it evaluates them based on their accumulated costs.

\subsection{Reinforcement learning models in behavioral economics}
\label{sec:learning}

In behavioral economics, reinforcement learning (RL) models belong to a broader class of stochastic learning models, which study how one or more players sequentially explore for optimal strategies in environments with unknown payoffs. In these models, players are assumed to adopt a stochastic policy (i.e., mixed strategies) and refine their strategies based on feedback. The roots of this line of research trace back to \citet{bush1955stochastic}, who conducted experiments on animal and human learning in unknown settings. To reproduce the observed behaviors, they proposed a simple RL rule in which the probability of choosing an action increases after positive reinforcement. 

While \citet{bush1955stochastic}'s model restricted outcomes to binary payoffs (positive or negative), many subsequent research efforts have generalized the framework to allow for continuous reward magnitudes. One example is \citet{cross1973stochastic}, who introduced a rule where actions that perform well increase their likelihood of being chosen again linearly with the reward. Interestingly, in the infinitesimal limit, this model aligns with replicator dynamics \citep{borgers1997learning}. Differently, \citet{arthur1991designing} proposed a model in which each action is assigned a ``strength" score that is updated after each round and then normalized to determine new choice probabilities, which captures the trade-off between exploiting successful actions and experimenting with others. Building on similar ideas, \citet{erev1998predicting} introduce internal ``propensities" for each option and use rewards to update those scores (the propensity of an option with high reward increases more), which then determine choice probabilities. They showed that such a simple model can explain many observed regularities in field experiments. As another example, \citet{camerer1999experience} developed the experience-weighted attraction model, which integrates reinforcement learning with belief-based approaches such as fictitious play \citep{monderer1996fictitious, berger2005fictitious}, thereby unifying two long-standing strands of learning theory in economics. For a comprehensive overview of how RL and related models have been applied and tested in behavioral economics, we refer readers to  \citet{camerer2011behavioral}. 
Our model fits squarely within this behavioral-economics RL class: the \emph{reinforcer} is the realized travel experience. The difference is that now the LLM supplies a qualitative assessment of which routes received positive reinforcement.

\subsection{LLM-driven strategic competition}
\label{sec:llm}

Recently, many researchers have begun to integrate LLMs into game-theoretic frameworks, aiming to examine whether LLM agents can act as rational players, how their strategies evolve in repeated interactions, and under what conditions their behavior converges to an equilibrium state.

Many initial research efforts focus on simple game-theoretic settings with few players. For example, \citet{guo2023gpt} examines GPT-4's behavior in the iterated Prisoner's Dilemma and shows that cooperation can be achieved when both players are prompted to value cooperation, whereas introducing even a single ``selfish" player largely disrupts cooperation and yields mutual defection. Extending beyond this, \citet{akata2025playing} find that in iterated Prisoner's Dilemma, GPT-4 is often ``unforgiving" and that in repeated Battle of the Sexes (a coordination problem where both players want to match actions but each prefers a different option), it rarely discovers the simple turn-taking convention (alternating between each player's preferred option). By contrast, \citet{meta2022human} find that equipping an LLM with explicit strategic planning and belief/opponent modeling yields reliable alliance formation in a multi-player negotiation game. 
Despite plausible strategic behavior, \textit{stable} outcomes remain difficult for prompt-only LLM agents, especially when equilibria require randomization (mixed strategies). For example, both \citet{silva2024large} and \citet{vidler2025playing} find that in Rock–Paper–Scissors (whose symmetric Nash equilibrium mixes $1/3$–$1/3$–$1/3$), LLMs may produce biased action frequencies rather than the equilibrium mixture. These results indicate that the randomized strategies required by mixed-strategy equilibria are hard for LLM agents to implement reliably without auxiliary mechanisms.

In game-theoretic scenarios that more closely approximate real-world scenarios, LLM agents have also shown potential for simulating human behavior. \citet{park2023generative} build an interactive ``small town" with 25 LLM agents and document believable social behaviors (e.g., organizing a Valentine's Day party). 
\citet{han2023guinea} run experiments with GPT-4–based firms in Bertrand-style price competition and find that without communication, the agents reach tacit collusion (prices converge above the Bertrand level but below monopoly), whereas with inter-firm messaging, they attain near-cartel prices more quickly. Yet, such demonstrations remain small-scale because of computational and cost constraints. For example, as highlighted by \citet{williams2023epidemic}, an LLM-based epidemic simulation with 1000 agents can require more than 90 hours of runtime on a 32GB RAM CPU machine. 

Also, many have explored LLM agents in transportation-related systems. \citet{wang2024can} study interactions among LLM-based autonomous-vehicle agents and find that social norms emerge and agents adopt conservative policies when facing potential collision risk. \citet{yan2026valuing} examine LLM-based values of travel time (VOT) and report sensitivity to income, trip purpose, and time--cost trade-offs. However, these studies also show that LLM behavior does not fully match human behavior, with weaker sensitivity to VOT and limitations in representing demographic differences \citep{yan2026valuing,voltesdorta2025airline}. Recent studies also investigate learning traveler personas from empirical choices \citep{liu2026persona} and calibrating persona-based synthetic mobility surveys to population statistics \citep{tzachristas2026pdm}. For day-to-day learning, \citet{liu2025llm} integrate LLM travelers into two scenarios: in a 40-day departure-time experiment at a single bottleneck, agents with short-term memory and bounded-rational updates approach the classical benchmark; in a 20-day two-route experiment, choices converge to the UE state with equal route travel times. \citet{wang2024ai} introduce an LLM-driven traveler with memory and day-to-day updating. Yet, even in a symmetric baseline with two identical routes, travelers keep changing routes day-to-day after a long time, so the system oscillates around equilibrium. Computational cost also raises concerns when extending these models to large-scale simulations. \citet{nishida2025modechoice}, for example, note that a single LLM-based mode-choice prediction takes approximately 0.75 seconds, compared with microseconds for conventional discrete-choice models. They identify the overhead of issuing large numbers of API requests, together with usage fees, as practical constraints on population-scale simulation.

This literature motivates two core design choices in our framework. First, to address scalability challenges in multi-agent LLM simulations, we replace many individual agents with representative agents. Second, to tackle the instability documented for prompt-only agents, especially when equilibria require mixed strategies, we separate \emph{reasoning} from \emph{updating}.

\section{The Proposed Framework}
\label{sec:framework}

We consider a standard nonatomic multi-class traffic assignment system. Let $\gW$ denote the set of origin--destination (OD) pairs, and let $\gV$ denote the set of traveler types. A composite class $m=(w,v)\in \gM=\gW\times\gV$ consists of travelers with OD pair $w$ and type $v$. Here, ``homogeneous'' means that travelers in the same composite class face the same perceived decision context and share similar decision-relevant profiles, such as income level, risk attitude, occupation, car ownership, or other attributes relevant to the application.
For each composite class $m \in \gM = \gW \times \gV$, let the total travel demand be $d_m > 0$, and let $\gK_m$ denote the set of commuting options available to travelers in that class, which may represent either different routes of the same mode or distinct modes. The flow profile on each day $t$ ($t = 0, 1, \ldots$) is then represented by $\vf_m^t = (\evf_{m,k}^t)_{k \in \gK_m} \in \sR_+^{|\gK_m|}$, which satisfies
$\sum_{k \in \gK_m} \evf_{m,k}^t = d_m$ for all $m \in \gM$.   On each day, the system's performance is then determined by the joint flow profile $\vf^t = (\vf_m^t)_{m \in \gM}$. For each class $m \in \gM$, let $e_{m,k}(\vf^t)$ denote the \textit{feedback} obtained from choosing commuting option $k \in \gK_m$. This feedback may include \textit{all relevant information} about the trip --- either numerical or semantic --- that influences travelers' preferences, rather than being restricted to a scalar cost.
For notational simplicity, we write $e_m(\vf^t) = (e_{m,k}(\vf^t))_{k \in \gK_m}$ for each $m \in \gM$ hereafter.
Under the described setting, our goal is to model how the aggregate flow $\vf_m^t$ of each $m \in \gM$ evolves over time until reaching a stable state.

\paragraph{Idea.}
To model the evolution of the system, rather than creating a large number of atomic travelers for each class $m$ as in conventional agent-based models, we assign a single nonatomic representative agent to each $m \in \gM$, whose strategy stands in for the entire class population. Each such agent is required to adopt a mixed strategy $\vp_m^t \in \Delta(\gK_m)$ on each day $t$, where
$\Delta(\gK_m) := \{ \vp_m \in \sR_+^{|\gK_m|} : \sum_{k \in \gK_m} \evp_{m,k} = 1\}.$
Assuming all travelers in class $m$ follow this common strategy, each traveler can be seen as independently sampling a commuting option from it each day. By the law of large numbers, the resulting aggregate flow is
$
\vf_m^t = d_m \cdot \vp_m^t, \quad m \in \gM, \quad t = 0, 1, \ldots.
$
Hence, modeling the evolution of the overall system dynamics reduces to characterizing how the strategies of the $|\gM|$ representative agents are updated over time. In the literature, there exist numerous well-established behavioral mechanisms for modeling how an agent rationally updates its mixed strategy $\vp^t$ based on daily experience, once each $e_{m,k}(\vf^t)$ is reduced to a scalar cost. In contrast, we aim to equip each agent with an LLM as a semantic evaluator, which allows them to interpret rich travel experiences with all relevant information and reason about how their strategies can be revised based on such information on their own. Before formalizing such ``representative LLM agents" --- as we shall refer to it --- we first provide a formal definition of LLM.

\begin{definition}
    \textit{An LLM is a parameterized stochastic conversational kernel $\Phi_\theta$ that generates a distribution over text responses conditioned on a preceding dialog history of messages. Each \emph{message} is a pair $\mu = (r, \, \omega)$, where $r$ denotes the \emph{role} and $\omega$ denotes the \emph{content}. In practice, the role can be: (1) \emph{system}, which provides overarching context and rules that the model is expected to follow throughout; (2) \emph{user}, which issues prompts or queries requiring an immediate response, and (3) \emph{agent}, which represents the model's replies. Given a dialog $\vh^t = (\mu^0, \ldots, \mu^{t-1})$, the LLM then generates the next message $(\text{\textit{agent}}, \, \omega^t)$, where $\omega^t \sim \Phi_\theta(\cdot \mid \vh^t)$.}
\end{definition}

With an equipped LLM $\Phi_{\theta}$, we can then interact with each agent using natural language. In the dialog, we use the system role to provide general instructions for the agent to follow before the simulation begins, and the user role to describe the travel experience and ask how the agent intends to adjust its strategy on each day. In the remainder of this section, we first outline how we communicate with each representative agent to prepare them for the simulation (Section~\ref{sec:initial}), before explaining how we interact with the agents through iterative exchanges, thereby modeling the resulting day-to-day dynamics (Section~\ref{sec:mechanism}).

\subsection{Initialization}
\label{sec:initial}

To prepare the agent for the subsequent interactions, we first need a system-role instruction prompt $\omega_m^{\text{s}}$, which delivers to the agent the information and guidelines it must keep in mind throughout the simulation.
This prompt should include three blocks: (1) $\omega_m^{\text{s-scenario}}$, which defines the agent as a traveler, endows it with relevant socio-demographic attributes, and provides a detailed clarification of all available commuting options between its origin and destination;
(2) $\omega_m^{\text{s-strategy}}$, a request to adopt and iteratively update a mixed strategy;
(3) $\omega_m^{\text{s-requirement}}$, the requirements governing its reasoning process and response format --- it depends on the types of questions that will be posed to the agent and will be discussed later. A general template of the system instruction is shown below.

\begin{system}[$\omega_m^{\text{s}} = (\omega_m^{\text{s-scenario}}, \omega_m^{\text{s-strategy}}, \omega_m^{\text{s-requirement}})$]
{\color{gray} $\omega_m^{\text{s-scenario}}:$} \textit{You are a daily commuter in a transportation network ... {\color{gray}(endow the agent with prescribed characteristics, if applicable)}. 
Each day, you can choose from $\{|\gK_m|\}$ commuting options, indexed by $1, \ldots, \{|\gK_m|\}$ ... {\color{gray}(describe each option with relevant details, such as its corresponding transport modes)}.}

{\color{gray} $\omega_m^{\text{s-strategy}}:$}  \textit{Your goal is to explore all available commuting options and eventually settle on one or more preferred choices. To do so, you will maintain a mixed strategy over these options and update it progressively according to your daily travel experience. }

{\color{gray} $\omega_m^{\text{s-requirement}}:$}  \textit{Each day, you will be asked to complete the following task(s) based on your current strategy and recent travel experiences ... {\color{gray}(describe the task(s))}. 
Your response must follow these requirements: ... {\color{gray}(specify reasoning guidelines and the required output format)}.}
\end{system}

Then, we engage the agent in a short dialog to establish an appropriate initial mixed strategy $\vp_m^0$, which can be either externally specified or determined by the agent itself. If chosen externally, we create the following user-role prompt $\omega_m^{t, \text{u-initial}}$, which informs the agent of its prescribed strategy:
\begin{user}[$\omega_m^{t, \text{u-initial}}$]
\textit{Suppose that your initial mixed strategy is $\{\vp_m^0\}$.}
\end{user}
In this case, the agent is not required to produce any response before the simulation begins; the dialog history can be directly initialized as
$
\vh_m^0 = [(\text{\textit{system}}, \omega_m^{\text{s}}),
(\text{\textit{user}}, \omega_m^{t, \text{u-initial}})].
$
Alternatively, the agent may determine its own initial strategy.
In this case, we construct the following user-role prompt $\omega_m^{t, \text{u-initial}}$, which directly asks the agent to formulate its initial mixed strategy:
\begin{user}[$\omega_m^{t, \text{u-initial}}$]
\textit{Based on your prior information about all commuting options, please think first, then provide the initial mixed strategy for exploration.
}
\end{user}
The initial strategy $\vp_m^0$ can then be extracted by querying the agent as
\begin{equation}
\omega_m^{t, \text{a-initial}} \sim
\Phi_\theta (\cdot \mid
[(\text{\textit{system}}, \omega_m^{\text{s}}), \;
(\text{\textit{user}}, \omega_m^{t, \text{u-initial}})]),
\label{eq:ask-initial}
\end{equation}
Accordingly, the dialog history before the simulation begins becomes
\begin{equation}
\vh_m^0 = [(\text{\textit{system}}, \omega_m^{\text{s}}), \;
(\text{\textit{user}}, \omega_m^{t, \text{u-initial}}), \;
(\text{\textit{agent}}, \omega_m^{t, \text{a-initial}})].
\label{eq:memory-initial}
\end{equation}
In the sequel, we assume that the initial strategy is determined by the agent itself unless otherwise stated. With both the initial strategy $\vp_m^0$ and dialog history $\vh_m^0$ established, the agent is then ready to engage in the subsequent day-to-day process through which both will evolve in response to the agent's daily experiences --- the underlying mechanism will be designed in the next section.

\subsection{Learning mechanisms}
\label{sec:mechanism}

On each day, the strategies adopted by the $|\gM|$ representative agents jointly determine the system's flow profile $\vf^t = (\vf_m^t)_{m \in \gM}$, which in turn gives rise to $e_m(\vf^t)$. This new experience is then conveyed to the agent through a natural-language prompt $\omega_m^{t,\text{u-feedback}}$, which explicitly describes $e_{m,k}(\vf^t)$ for each $k \in \gK_m$. Upon receiving this information, the agent interprets it and updates its strategy accordingly. Formally, this learning process can be represented as a mapping
$
\pi_m: (\vp_m^t, \vh_m^t; \omega_m^{t,\text{u-feedback}}) \mapsto (\vp_m^{t+1}, \vh_m^{t+1}),
$
which defines how both the agent's strategy and the dialog history evolve in response to the newly received feedback. Given $\pi_m$, the dynamics can then be modeled via Algorithm~\ref{alg:overview}.

\begin{algorithm}[ht]
   \caption{Representative-LLM framework for modeling day-to-day route choice}
   \label{alg:overview}
   \makeatletter
\renewcommand{\ALG@beginalgorithmic}{\setlength{\baselineskip}{0.8\baselineskip}}
\makeatother
   \small
   \begin{algorithmic}[1]
   \For {each representative agent $m \in \gM$}
      \State Query the agent to obtain $\omega_m^{t, \text{a-initial}}$ per Equation \eqref{eq:ask-initial} to extract $\vp_m^0$, and then initialize  $\vh_m^0$ per Equation \eqref{eq:memory-initial}.
   \EndFor
   \For {$t = 0, 1, \ldots$}
      \State Compute $\vf_m^t = d_m \cdot \vp_m^t$ for all $m \in \gM$ and then set $\vf^t = (\vf_m^t)_{m \in \gM}$.
      \For {each representative agent $m \in \gM$}
         \State Generate a prompt $\omega_m^{t,\text{u-feedback}}$ describing $e_{m}(\vf^t)$, and then set
         $
         (\vp_m^{t+1}, \vh_m^{t+1}) = \pi_m(\vp_m^t, \vh_m^t;\, \omega_m^{t,\text{u-feedback}}).
         $
         
      \EndFor
   \EndFor
\end{algorithmic}
\end{algorithm}

We now turn to how the power of LLMs can be leveraged to design the learning mechanism $\pi_m$. We first present two approaches in which the strategy updates are performed entirely by the LLM itself (Sections \ref{sec:full-llm} and \ref{sec:full-llm2}), before motivating an LLM-guided mechanism (Section \ref{sec:llm-guided}), which integrates the LLM's reasoning with a reinforcement-learning update rule.

\subsubsection{A fully LLM-driven baseline learning mechanism} 
\label{sec:full-llm}

As a baseline, we let the LLM take full responsibility for revising the agent's strategy.
To this end, we construct a user-role prompt $\omega_m^{t,\text{u-new}}$ that directly requests the agent to produce a revised mixed strategy for the next day.
An example prompt is shown below:
\begin{user}[$\omega_m^{t,\text{u-new}}$]
    \textit{Please revise your strategy by reflecting on your current strategy and your travel experiences in recent days.}
\end{user}
The agent's response can then be queried as
\begin{equation}
\omega_m^{t, \text{a-new}} \sim
\Phi_\theta(\cdot \mid
[\vh_m^t, \; (\text{\textit{user}}, \omega_m^{t,\text{u-feedback}}), \;(\text{\textit{user}}, \omega_m^{t,\text{u-new}})]).
\label{eq:baseline-query}
\end{equation}
By parsing the agent's response $\omega_m^{t, \text{a-new}}$, we extract the updated strategy $\vp_m^{t+1}$.
The dialog history is then extended to include this daily exchange.
Formally, the process is summarized in Algorithm~\ref{alg:baseline}.

\begin{algorithm}[ht]
   \caption{A baseline learning mechanism $\pi_m: (\vp_m^t, \vh_m^t; \omega_m^{t, \text{u-feedback}}) \mapsto (\vp_m^{t + 1}, \vh_m^{t + 1})$}
   \label{alg:baseline}
   \makeatletter
\renewcommand{\ALG@beginalgorithmic}{\setlength{\baselineskip}{0.8\baselineskip}}
\makeatother
   \small
\begin{algorithmic}[1]
   
   \State Query the agent to obtain $\omega_m^{t, \text{a-new}}$ per Equation \eqref{eq:baseline-query}.
   \State Parse $\omega_m^{t,\text{a-new}}$ to extract the revised mixed strategy $\vp_m^{t+1}$ and set
   \begin{equation*}
   \vh_m^{t+1} = [\vh_m^t, (\text{\textit{user}}, \, \omega_m^{t,\text{u-feedback}}), \, (\text{\textit{user}}, \, \omega_m^{t,\text{u-new}}), \, (\text{\textit{agent}}, \, \omega_m^{t, \text{a-new}})].
   \end{equation*}
\end{algorithmic}
\end{algorithm}

This illustrates the most direct way to use LLMs for adaptive learning. While straightforward, the entire update process is left to the black-box behavior of the LLM, making it difficult to interpret or control how the strategy evolves. We have no clear understanding of the reasoning behind the agent's revisions, nor any guarantee that a consistent logic will be followed across iterations.

\subsubsection{A fully LLM-driven RL mechanism} 
\label{sec:full-llm2}

To improve interpretability, we propose to decompose the update process into two stages --- in the spirit of the law of effect \citep{thorndike1927law}  --- following a chain-of-thought rationale.

First, the LLM is asked to reason about its recent travel experiences and identify which commuting options should be \emph{positively reinforced} --- that is, used more frequently in the future; then, a subsequent prompt requests the agent to think of an updated mixed strategy.
First, the LLM is prompted to qualitatively interpret the feedback and identify a subset of actions $\gK_m^+ \subset \gK_m$ that it views as positively reinforced, that is, routes the traveler would prefer to use more often after the current experience.
An example of this prompt, denoted as $\omega_m^{t, \text{u-positive}}$, is shown below.
\begin{user}[$\omega_m^{t, \text{u-positive}}$]
   \textit{Please reflect on your current strategy and your travel experiences in recent days, and then select a subset of the commuting options for which you would like to increase the selection probability for the next day.}
\end{user}
The new prompt allows us to query the agent as follows:
   \begin{equation}
     \omega_m^{t,\text{a-positive}}
     \sim \Phi_{\theta}(\cdot \mid
       [\vh_m^t,\;
       (\text{\textit{user}},  \omega_m^{t,\text{u-feedback}}),\;
       (\text{\textit{user}}, \; \omega_m^{t,\text{u-positive}})]).
       \label{eq:llm-rl-1}
   \end{equation}
The next question then follows up to ask the agent how exactly it would like to update the strategy:
\begin{user}[$\omega_m^{t, \text{u-revise}}$]
    \textit{Please indicate how you will update your strategy to reinforce the use of these options.}
\end{user}
The agent is then queried to provide its revised strategy in response to this follow-up prompt:
\begin{equation}
\omega_m^{t,\text{a-revise}}
\sim \Phi_{\theta}(\cdot \mid
[\vh_m^t, \,
(\text{\textit{user}}, \omega_m^{t,\text{u-feedback}}), \,
(\text{\textit{user}}, \omega_m^{t,\text{u-positive}}), \,
(\text{\textit{agent}}, \omega_m^{t,\text{a-positive}}), \,
(\text{\textit{user}}, \omega_m^{t,\text{u-revise}})]).
\label{eq:llm-rl-2}
\end{equation}
By parsing the agent's response $\omega_m^{t,\text{a-revise}}$, the updated strategy $\vp_m^{t+1}$ can be extracted, and the dialog history is accordingly expanded to record this two-round conversation before proceeding to the next day. Formally, the above learning process is summarized in Algorithm \ref{alg:baseline2}.

\begin{algorithm}[ht]
   \caption{A fully LLM-driven RL mechanism $\pi_m: (\vp_m^t, \vh_m^t; \omega_m^{t, \text{u-feedback}}) \mapsto (\vp_m^{t + 1}, \vh_m^{t + 1})$}
   \label{alg:baseline2}
   \makeatletter
\renewcommand{\ALG@beginalgorithmic}{\setlength{\baselineskip}{0.8\baselineskip}}
\makeatother
   \small
\begin{algorithmic}[1]
   
   \State Query the agent to obtain $\omega_m^{t, \text{a-positive}}$ per Equation \eqref{eq:llm-rl-1}.
   \State Query the agent to obtain $\omega_m^{t, \text{a-revise}}$ per Equation \eqref{eq:llm-rl-2}.
   \State Parse $\omega_m^{t,\text{a-revise}}$ to extract the revised mixed strategy $\vp_m^{t+1}$ and set
   \begin{equation*}
     \vh_m^{t+1} = [\,\vh_m^t,\; (\text{\textit{user}}, \, \omega_m^{t,\text{u-feedback}}),\; 
     (\text{\textit{user}}, \, \omega_m^{t,\text{u-positive}}),\;
     (\text{\textit{agent}}, \, \omega_m^{t,\text{a-positive}}),\;
    (\text{\textit{user}}, \, \omega_m^{t, \text{u-revise}}),\; (\text{\textit{agent}}, \, \omega_m^{t, \text{a-revise}})\,].
   \end{equation*}

\end{algorithmic}
\end{algorithm}

Compared with the previous approach, the new one improves interpretability by inducing a chain of thought in the agent's reasoning process so that the update of its mixed strategy now follows a more traceable logic. However, the dynamics remain fully LLM-driven: the model autonomously determines how strategies evolve from day to day without any external control or stabilization mechanism. Consequently, there is no guarantee that the learning process will converge, nor that the fluctuations in the resulting flows will diminish over time.

\subsubsection{An LLM-guided RL mechanism}
\label{sec:llm-guided}

To enhance both interpretability and stability, we propose a new approach that retains the RL structure but separates reasoning from updating: the LLM provides qualitative judgments on which actions should be reinforced, while the quantitative update of the mixed strategy is governed by explicit, rule-based mechanisms. The first step, as in the previous mechanism, is to query the agent per Equation~\eqref{eq:llm-rl-1}, so that by parsing the response $\omega_m^{t,\text{a-positive}}$, we can extract the positively reinforced subset of commuting options $\gK_m^+ \subset \gK_m$.
The second step, however, is no longer delegated to the agent itself. If the agent returns an empty set $\gK_m^+$ (i.e., it indicates that no options should be reinforced), no further adjustment is needed, and we directly proceed to the next day. Otherwise, when $\gK_m^+ \neq \emptyset$, the agent's mixed strategy is updated according to a predefined rule $\psi_{m, \, \gK_m^+}: \Delta(\gK_m) \to \Delta(\gK_m)$, which systematically shifts probability mass toward the positively reinforced actions.
Specifically, we propose two such update rules in what follows.
\begin{itemize}[leftmargin=*, topsep=0pt, itemsep=2pt, parsep=0pt]
    \item \textbf{Rule 1}: A simple example for $\psi_{m, \, \gK_m^+}: \vp_m^t \mapsto \vp_m^{t + 1}$ is given by, for each $k \in \gK_m$,
    \begin{equation}
    p_{m,k}^{t+1} =
    \begin{cases}
    \dfrac{s_m^{t+1}}{s_m^t} \cdot p_{m,k}^t, & \text{if } k \in \gK_m^+, \\[5pt]
    (1 - \eta_m^t) \cdot p_{m,k}^t, & \text{if } k \notin \gK_m^+,
    \end{cases}
    \quad \text{where}~ 
    \begin{cases}
    s_m^t = \sum_{k \in \gK_m^+} p_{m,k}^t \\[5pt]
    s_m^{t+1} = (1 - \eta_m^t)\cdot s_m^t + \eta_m^t \cdot 1,
    \end{cases}
    \ \text{and} \ \eta_m^t \in (0,1).
    \label{eq:rule1}
    \end{equation}
    This rule, similar to \citet{bush1955stochastic}'s RL model, proportionally increases the probabilities of actions in $\gK_m^+$ while shrinking all others to preserve normalization.
    \item \textbf{Rule 2}: Another example is inspired by the multiplicative weights update method \citep{arora2012multiplicative}, which assumes, for each $k \in \gK_m$,
    \begin{equation}
        p_{m,k}^{t+1} 
        = \frac{p_{m,k}^t \cdot \exp( \sigma_{m, k} \cdot \eta_m^t)}
        {\sum_{j \in \gK_m} p_{m,j}^t \cdot \exp(\sigma_{m, j} \cdot \eta_m^t)},
        \quad \text{where}~
        \sigma_{m, j} = 
        \begin{cases}
        1, & j\in \gK_m^+, \\[-2pt]
        0, & j \notin \gK_m^+
        \end{cases}
        ~~\text{and}~~\eta_m^t \in (0, 1).
        \label{eq:rule2}
    \end{equation} 
\end{itemize}
Either way, the larger $\eta_m^t$ is, the more aggressively new $\vp_m^{t + 1}$ shifts probability mass towards $\gK_m^+$.

The next challenge is how to extend the dialog history. Although the strategy update in this mechanism is executed externally rather than through an actual conversation with the agent, it must still be documented to preserve continuity and provide a complete memory for the next day's interaction. A straightforward approach would be to insert a user-role prompt that directly informs the agent that its strategy has been updated to $\vp_m^{t+1}$. However, this is semantically unnatural because we have allowed the agent to exercise autonomy in identifying which commuting options it prefers to reinforce. To maintain a more natural flow of the dialog, we instead ``simulate'' a short conversation to be appended to the dialog, which consists of a user-side instruction $\omega_m^{t, \text{u-revise}}$ requesting the update (the same prompt used in Algorithm~\ref{alg:baseline2}) followed by an agent-side confirmation $\omega_m^{t, \text{a-revise}}$ that states the updated strategy (not actually generated by the LLM), as shown below:
\begin{agent}[$\omega_m^{t, \text{a-revise}}$]
\textit{I would like to update my mixed strategy to {$\vp_m^{t+1}$} for tomorrow's use.}
\end{agent}

In summary, this \emph{LLM-guided RL} mechanism, where the LLM only guides a direction of learning while the update follows an explicit rule, is formalized as Algorithm~\ref{alg:rl}.
In contrast to the previous two approaches, the proposed mechanism makes stability explicitly tunable: by scheduling $\eta_m^t \to 0$, the dynamics are forced to ``cool," ensuring that adjustments gradually vanish and $\|\vp_m^{t+1}-\vp_m^t\| \to 0$. { We suggest a decay rate on the order of $\gO(1/t)$, motivated by classical assignment algorithms such as the method of successive averages (MSA). In the scalar-cost setting, MSA combines an all-or-nothing direction toward the current minimum-cost routes with a diminishing step size, and convergence to UE is established under standard assumptions when an appropriate decaying schedule, classically of order $\gO(1/t)$, is used. Here, if the step size decays too slowly, the algorithm may diverge; if it decays too quickly, learning may freeze before reaching UE. We therefore adopt the same order of decay in the proposed LLM-guided dynamics. However, because the LLM may reinforce multiple competitive routes rather than only the current minimum-cost route, this step-size choice should be viewed as a stabilizing heuristic rather than a theoretical guarantee of convergence.}

\begin{algorithm}[ht]
   \caption{An LLM-guided RL module $\pi_m: (\vp_m^t, \vh_m^t; \omega_m^{t, \text{u-feedback}}) \mapsto (\vp_m^{t + 1}, \vh_m^{t + 1})$}
   \label{alg:rl}
   \makeatletter
\renewcommand{\ALG@beginalgorithmic}{\setlength{\baselineskip}{0.8\baselineskip}}
\makeatother
   \small
\begin{algorithmic}[1]
   \State Query the agent to obtain $\omega_m^{t, \text{a-positive}}$ per Equation \eqref{eq:llm-rl-1}.
    \State Parse $\omega_m^{t,\text{a-positive}}$ to extract a positively reinforced subset of actions $\gK_m^+ \subset \gK_m$.
    \State \textbf{If} $\gK_m^+=\emptyset$, \textbf{then} set $\vp_m^{t+1} = \vp_m^t$ and
       \begin{equation*}
       \vh_m^{t+1} = [\vh_m^t, \,(\text{\textit{user}}, \, \omega_m^{t,\text{u-feedback}}), \,(\text{\textit{user}}, \, \omega_m^{t,\text{u-positive}}),  (\text{\textit{agent}}, \, \omega_m^{t, \text{a-positive}})].
       \end{equation*}
    \State \textbf{If} $\gK_m^+ \neq \emptyset$, \textbf{then} compute $\vp_m^{t+1} = \psi_{m, \, \gK_m^+}(\vp_m^t)$ via either Equations \eqref{eq:rule1} or \eqref{eq:rule2}
    and then set
   \begin{equation*}
     \vh_m^{t+1} = [\,\vh_m^t,\, (\text{\textit{user}}, \, \omega_m^{t,\text{u-feedback}}),\, 
     (\text{\textit{user}}, \, \omega_m^{t,\text{u-positive}}),\,
     (\text{\textit{agent}}, \, \omega_m^{t,\text{a-positive}}),\,
    (\text{\textit{user}}, \, \omega_m^{t, \text{u-revise}}),\, (\text{\textit{agent}}, \, \omega_m^{t, \text{a-revise}})\,].
   \end{equation*}
\end{algorithmic}
\end{algorithm}

\begin{remark}[Requirement block]
\label{rm:structured}

As noted in Section~\ref{sec:initial}, to prevent overly freestyle outputs, the system-role prompt $\omega_m^{\text{s}}$ should include a requirement $\omega_m^{\text{s-requirement}}$ to make the LLM’s daily response structured and parsable. This block specifies the task to be completed on each day, the expected reasoning style, and the exact output format. The mechanism-level requirement blocks for Algorithms~\ref{alg:baseline}, ~\ref{alg:baseline2}, and ~\ref{alg:rl} follow the same principle, with minor changes matching their different interaction flows. The full prompt designs are provided in Appendix~\ref{app:additional}.

\end{remark}

{
\begin{remark}[Repeated LLM queries and voting]
\label{rm:voting}
The LLM-guided RL mechanism in Algorithm 4 uses the LLM to identify the positively reinforced set $K_m^+$. Since the LLM is a stochastic generative model, a single query may occasionally produce an ``idiosyncratic" reinforcement decision. For applications where a stable estimate of the representative agent's central tendency is desired, one may replace the single query in Step 1 of Algorithm 4 with several independent queries under the same dialog history and feedback prompt. The final reinforced set $K_m^{+,t}$ is then determined by majority voting, or equivalently by selecting the most frequently returned candidate set.    
\end{remark}
}

{
\begin{remark}[Conversation history compression]
\label{rm:memory}
In the current description, the dialog history $\vh_m^{t}$ appends the interaction from all previous days. This preserves continuity in short-horizon simulations. However, for long-horizon applications, full-history prompting may approach the LLM's context-window limit and increase token usage over time. Moreover, it is not necessary to assume that the agent relies on all past experiences equally: recent feedback is typically more salient for day-to-day adjustment. Therefore, one may replace full-history prompting with bounded memory, such as keeping recent days verbatim while summarizing earlier experience or just simply retaining only the most recent $H$ days (e.g., $H = 5$).
\end{remark}
}

\section{Application to Classic Traffic Assignment}
\label{sec:application-tap}

We first apply the proposed framework to the classic traffic assignment problem. In such settings, UE is the established benchmark. If the resulting dynamics were unstable or deviated substantially from UE in this basic scalar-cost setting, its use in richer behavioral settings would be less convincing. This experiment also provides a direct comparison between the proposed scheme and baselines where the LLM is responsible for revising the mixed strategy itself.

Unless otherwise specified, all experiments use the Python API of DeepSeek-V3.1 for LLM interactions. Temperature controls the randomness of token sampling, while ``top\_p'' restricts the candidate tokens based on cumulative probability. Following DeepSeek's recommendation to adjust only one of these parameters, we keep ``top\_p'' at its default value of 1 and set temperature to 0.2 \citep{deepseek_api_sampling}. This low but nonzero temperature is intended to limit unnecessary sampling variation while retaining some response diversity. For the LLM-guided RL mechanism, the learning rate is specified as $\eta_m^t=c/(b+t)$, with $c=1$ and $b=3$.

\subsection{Problem setting}

A classic traffic assignment problem is defined on a directed graph $\gG(\gN, \gA)$, where $\gN$ and $\gA$ denote the sets of nodes and links, respectively. Travelers between all origin–destination (OD) pairs $\gW \subseteq \gN \times \gN$ are assumed to be homogeneous. For each OD pair $w \in \gW$, denote $\gK_m$ as all feasible routes $\gR_w \subseteq 2^{\gA}$ connecting $w$, and $\gA_{w,r} \subseteq \gA$ as the set of links used by each route $r \in \gR_w$. Letting $f_{w,r}$ be the number of travelers choosing route $r \in \gR_w$ for OD pair $w$, and $x_a$ be the flow on link $a \in \gA$, it must then hold that $x_a = \sum_{w \in \gW} \sum_{r \in \gR_w} \emLambda_{a,r}^w \cdot f_{w, r}$ for all $a \in \gA$,
where $\Lambda_{a,r}^w = 1$ if $a \in \gA_{w,r}$ and $0$ otherwise.
Assume each link $a \in \gA$ has a continuous, increasing travel time function $u_a(\evx_a)$. For notational convenience, write the aggregate route flow as $\vf = (\vf_w)_{w \in \gW}$, where $\vf_w = (f_{w,r})_{r \in \gR_w}$, and the feasible region for $\vf$ as $\gF$. Then, for any OD pair $w \in \gW$, the route cost for $r \in \gR_w$ can be defined as a function $c_{w,r}: \gF \to \sR$ such that
\begin{equation}
c_{w,r}(\vf) = \sum_{a \in \gA} \Lambda_{a,r}^w \, t_a(x_a),
\quad \text{with} \quad
x_a = \sum_{w \in \gW} \sum_{r \in \gR_w} \Lambda_{a,r}^w \, f_{w,r}.
\end{equation}
For notational simplicity, we write $c(\vf) = (c_w(\vf))_{w \in \gW}$, where $c_w(\vf) = (c_{w, r}(\vf))_{r \in \gR_w}$, hereafter.

In conventional traffic assignment, it is typically assumed that travelers aim to minimize their travel time, which will drive the system to the UE state where no traveler can reduce their travel cost by unilaterally switching their routes \citep{wardrop1952road}. Formally, a feasible route flow pattern $\vf \in \gF$ satisfies the UE condition if for each OD pair $w \in \gW$, there exists a constant $b_w \in \sR$ such that for all $r \in \gR_w$, it holds that $c_{w,r}(\vf) = b_w$ if $f_{w,r} > 0$, and $c_{w,r}(\vf) \geq b_w$ if $f_{w,r} = 0$. 
To measure the deviation of a flow pattern $\vf$ from UE, we use the relative gap:
\begin{equation}
\delta(\vf)
= -\frac{\langle c(\vf), \vf' - \vf \rangle}{\langle c(\vf), \vf \rangle},
\quad \text{where}~
\vf' = \argmin_{\vf'' \in \gF} , \langle c(\vf), \vf'' \rangle.
\end{equation}
It is straightforward to verify that $\delta(\vf) = 0$ if and only if $\vf$ satisfies the UE condition.

\subsubsection{Implementation and results}

The proposed framework will be evaluated on three benchmark networks commonly used in the transportation literature: a 3-node-4-link (3N4L) network \citep{tobin1988sensitivity}, the Hearn network, and the Sioux-Falls network; full details are available in the Transportation Networks GitHub Repository \citep{TransportationNetworks}. The 3N4L network consists of 3 nodes, 4 links, and a single OD pair. The Hearn network contains 9 nodes, 18 links, and 4 OD pairs. The Sioux-Falls network has 24 nodes, 76 links, and 528 OD pairs. In the Sioux-Falls network, a subset of routes, which covers all UE routes as well as many non-UE routes, is pre-selected for simulation.

As all travelers are assumed to be homogeneous, $\gM = \gW$, and for each $m \in \gM$, the set $\gK_m$ simply represents all feasible routes $\gR_w \subseteq 2^{\gA}$ connecting the corresponding OD pair $w$. Accordingly, to implement the proposed framework, it suffices to create $|\gW|$ representative LLM agents, one for each OD pair. For each of these agents, the scenario-dependent component $\omega_m^{\text{s-scenario}}$ of $\omega_m^{\text{s}}$ informs the agent that it is a daily commuter in a transportation network, that it may choose among the feasible routes indexed in $\gK_m$. Meanwhile, since travelers aim purely to minimize travel time, the feedback prompt $\omega_m^{t,\text{u-feedback}}$ therefore simply describes the travel times on all routes between the corresponding OD pair. The exact formats of both the scenario prompt and feedback prompt are provided in Appendix~\ref{app:prompt-classic}.

The three learning mechanisms introduced in Algorithms~\ref{alg:baseline}--\ref{alg:rl} are all tested and compared. Using each mechanism, the day-to-day dynamics (Algorithm~\ref{alg:overview}) are simulated over 50 days. For the LLM-guided RL mechanism (Algorithm~\ref{alg:rl}), both adjustment rules defined in Equations~\eqref{eq:rule1} and~\eqref{eq:rule2} are examined, with the learning rate $\eta^t$ decaying at an $\mathcal{O}(1/t)$ rate. { For benchmarking, we also implement a classical best-response/MSA dynamics, where for each OD pair and each day, the LLM-generated positively reinforced set is replaced by the route with the lowest realized cost from the previous day, and the mixed strategy is updated using the same $\mathcal{O}(1/t)$ step-size schedule. This benchmark is included to provide a familiar equilibrium-seeking reference trajectory against which the stability of the LLM-based dynamics can be assessed.}

\begin{figure}[ht]
    \centering
    \includegraphics[width=0.95\textwidth]{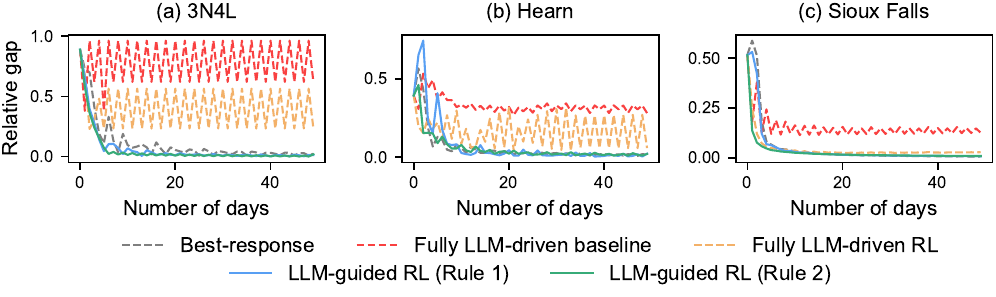}
    \captionof{figure}{Evolution of the relative equilibrium gap under different learning mechanisms across three networks.}
    \label{fig:exp-1}
\end{figure}

Figure \ref{fig:exp-1} shows the evolution of the relative equilibrium gap over simulated days under different learning mechanisms across three benchmark networks.
The two fully LLM-driven baselines exhibit strong oscillations and fail to converge: large day-to-day fluctuations persist even after 50 iterations, with the relative gap remaining well above zero; the only exception is the fully LLM-driven RL dynamics on the Sioux Falls network, where the gap quickly drops to around 0.02 and then stabilizes. This behavior arises because the LLM agent, acting autonomously, does not ``slow down" its learning over time --- that is, it lacks an inherent mechanism for step-size decay --- although such temporal information is implicitly present in the dialogue history and could be leveraged.
In contrast, the proposed LLM-guided RL dynamics (both Rule 1 and Rule 2) converge rapidly and stably toward UE across all networks. Their convergence speed and stability are comparable to those of the best-response dynamics, showing only negligible residual fluctuations after about ten iterations.

We further examine the LLM agents' responses to better understand their reinforcement patterns; see Appendix~\ref{app:classic-response} for details. In brief, we identify three patterns that differ from a simple best-response rule: (1) the agent may reinforce not only the shortest route, but also another route that is close in cost and has recently improved; (2) the agent may decline to further reinforce the shortest route that already has a very high selection probability; (3) the agent may treat a very small cost difference as negligible and choose no reinforcement. These examples suggest that the LLM can use qualitative reasoning to determine when multiple routes are worth reinforcing, when additional reinforcement is unnecessary, and when a small cost difference can be neglected.

In conclusion, this experiment highlights the importance of separating reasoning from updating. Even in this simple scalar-cost setting, fully delegating the mixed-strategy update to the LLM leads to unstable dynamics that do not reliably approach UE. By contrast, the proposed LLM-guided RL mechanism combines the LLM's qualitative ability to identify routes worth reinforcing with an explicit decaying-step-size update, which yields stable dynamics that approach near-UE states.

\section{Applications to Richer Traffic Settings}
\label{sec:application-rich}

We then examine the framework in three behaviorally richer traffic settings, which are designed as proof-of-concept applications in which travelers must reason over multiple attributes, contextual comparisons, or profile-dependent preferences, rather than only scalar travel time. Specifically, we consider: (1) a highway tolling scenario, where travelers balance travel time and monetary tolls and may respond to the introduction of an inferior toll route (Section~\ref{sec:tolling}); { (2) a transit transfer scenario with denied-boarding risk, where travelers trade off a riskier transfer option against a more reliable direct option (Section~\ref{sec:risk});} (3) a multi-modal commuting scenario, where travelers from different income groups choose among public transit, driving, and park-and-ride options with different time, cost, comfort, and convenience attributes (Section~\ref{sec:multi-modal}).

Hereafter, we focus only on testing the proposed LLM-guided RL scheme (Algorithm~\ref{alg:rl}), as the two baselines cannot remain stable even in the simplest classic traffic assignment settings. 
Unless otherwise specified, only Rule 1 is employed, as both rules produced comparable outcomes in the previous experiment. The LLM sampling configuration and learning-rate decay schedule are the same as in the previous experiment. 
Unlike the classic traffic assignment experiment, however, where the dynamics can be compared against a unique UE link-flow benchmark, the richer scenarios studied below may exhibit non-negligible run-to-run variation due to the stochastic LLM component. Therefore, each setting is repeated 10 times. In the subsequent results, we retain all 10 runs to make the full range of LLM-induced variation visible. At the same time, to provide a stable central-tendency summary for behavioral interpretation, we compute a ``trimmed" trajectory summary: the six runs closest to the full-run mean trajectory are selected, and their mean is used as the central trajectory.

\subsection{Highway tolling}
\label{sec:tolling}

The experiment is conducted on a simple highway network with one OD pair connected by three parallel routes: a non-toll road (Road 1) and two toll roads (Roads 2 and 3). All travelers are homogeneous, with a monthly salary of 25,000 HKD, and a total travel demand of $d = 10$. The travel time on each road (in minutes) follows the Bureau of Public Roads (BPR) form $u_{\text{time}, a}(\evx_a) = \evu_{0, a} \cdot (1 + 0.15 \cdot (\evx_a / \evv_{a})^{4})$ ($a = 1, 2, 3$), where $\evu_{0, a}$ denotes the free-flow travel time and $\evv_a$ the practical capacity. We consider three parameter settings. In Setting A, $(\evu_{0,1},\evv_1,\pi_1)=(45,8,0)$, $(\evu_{0,2},\evv_2,\pi_2)=(30,8,30)$, and $(\evu_{0,3},\evv_3,\pi_3)=(37,8,34)$. Setting B is the same as Setting A except that the capacity of Road 2 is reduced to $\evv_2=4$. Setting C is the same as Setting A except that the free-flow time of Road 1 is reduced to $\evu_{0,1}=40$. Under each setting, we compare two scenarios --- without and with Road 3 --- to examine how introducing an additional toll road influences travelers' route choices.

\subsubsection{Implementation}

Since the network consists of a single OD pair and all travelers are assumed homogeneous, we have $|\gM| = 1$, and $\gK_m$ for the sole class $m$ corresponds to the available routes. To implement the proposed framework, only one representative agent needs to be created. The scenario-dependent component $\omega_m^{\text{s-scenario}}$ of the system-role prompt is designed to specify the agent's socioeconomic profile and briefly introduces the commuting routes available for selection. For this experiment, an additional instruction is appended to the requirement block to guide the agent toward boundedly rational, human-like reasoning; the rationale is discussed in Appendix~\ref{app:prompt-tolling}. The feedback $e_{m,k}(\vf^t)$ for each route on each day now includes two components: the realized travel time and the monetary toll. The exact formats of the scenario prompt, additional instruction, and feedback prompt are provided in Appendix~\ref{app:prompt-tolling}.

To compare the proposed LLM-guided framework with a standard scalar-cost benchmark, we also implement a UE model based on generalized travel cost. In conventional tolling models, time and money are typically combined through a value-of-time parameter: for each route \(a\), the generalized cost is written as $c_a(\evx_a)=\lambda \cdot u_{\text{time},a}(\evx_a)+\pi_a$, where $\lambda$ denotes the value of time in HKD per minute. The equilibrium outcome is then predicted by applying the UE condition to these generalized costs. In each setting (A, B, and C), we first calibrate $\lambda$ so that, in the two-route scenario without Road 3, the UE model reproduces the long-run route shares generated by the LLM-guided dynamics. We then hold the same calibrated $\lambda$ fixed and compute the UE prediction for the three-route scenario with Road 3. This design gives the scalar-cost UE benchmark its best match to the LLM-guided outcome in the baseline two-route case, so that any difference in the three-route case reflects how the two approaches respond to the introduction of the additional toll road.

\subsubsection{Results}

\begin{figure}[ht]
    \centering
    \includegraphics[width=0.9\textwidth]{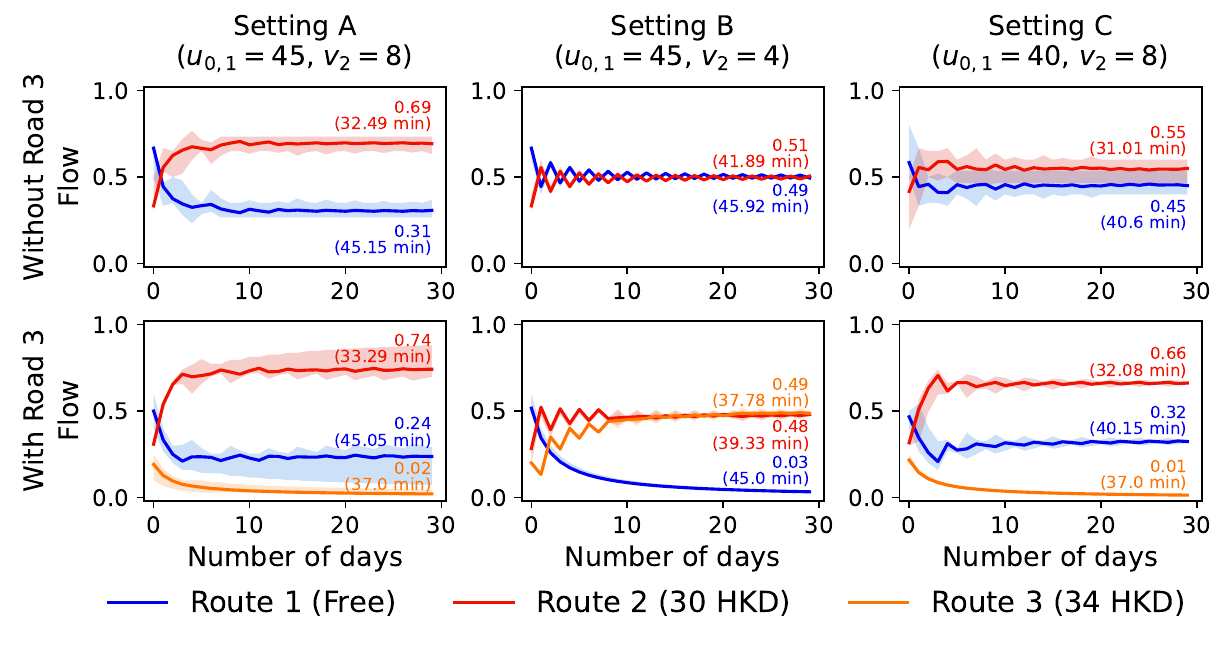}
    \captionof{figure}{Evolution of route flows under the LLM-guided RL dynamics for the three highway-tolling settings (A--C), comparing scenarios with and without Road 3. { Solid lines denote trimmed central-tendency trajectories; darker and lighter shaded areas denote the six-run and ten-run spreads, respectively.}}
    \label{fig:exp2-result}
\end{figure}

Figure~\ref{fig:exp2-result} illustrates how the proportions of travelers selecting each route evolve under the proposed LLM-guided RL dynamics, whereas Table~\ref{tab:ue_llm_comparison} compares the LLM-guided outcomes with the calibrated scalar-cost UE benchmark (the LLM-guided row in the table reports the trimmed central-tendency final shares). It can first be seen from Figure~\ref{fig:exp2-result} that the LLM-guided dynamics stabilize quickly across all three settings, typically within about 10--20 iterations. 

\begin{table}[ht]
\centering
\caption{Comparison of LLM-guided and UE-based predictions under the scenario \textit{with} Road 3 (the UE model is calibrated to reproduce the LLM outcomes obtained in the scenario \textit{without} Road 3).}
\label{tab:ue_llm_comparison}
\setlength{\tabcolsep}{6pt}
\small
\begin{tabular}{l ccc ccc ccc}
\toprule
\multirow{2}{*}{\textbf{Method}} 
& \multicolumn{3}{c}{\textbf{Setting A}} 
& \multicolumn{3}{c}{\textbf{Setting B}} 
& \multicolumn{3}{c}{\textbf{Setting C}} \\
\cmidrule(lr){2-4} \cmidrule(lr){5-7} \cmidrule(lr){8-10}
 & Route 1 & Route 2 & Route 3 
 & Route 1 & Route 2 & Route 3 
 & Route 1 & Route 2 & Route 3 \\
\midrule
UE-based  & 31\% & 69\% & 0\%  & 0\% & 47\% & 53\% & 45\% & 55\% & 0\% \\
LLM-based & 24\% & 74\% & 2\%  & 3\% & 48\% & 49\% & 32\% & 66\% & 1\% \\
\bottomrule
\end{tabular}
\end{table}

Under Setting B, repeating the LLM-guided dynamics produces nearly identical outcomes, even though LLM outputs are typically stochastic. This stability appears to arise because the available alternatives have clear trade-offs. Without Road 3, the faster toll road and the slower toll-free road are both attractive for different reasons. After Road 3 is introduced, the two tolled routes become close substitutes: Road 2 is slightly cheaper, whereas Road 3 is slightly faster. As a result, the LLM-guided outcome assigns comparable shares to the two tolled routes, which is broadly consistent with the calibrated UE-based prediction in Table~\ref{tab:ue_llm_comparison}.

Under Settings A and C, however, the LLM-guided and UE-based predictions differ in an important way. Under Setting A, when Road 3 is absent, approximately 31\% of travelers choose the slower, toll-free route, while about 69\% are willing to pay 30 HKD to save 13 minutes by taking the toll road. When the slightly more expensive Road 3 is introduced, its long-run travel time is roughly four minutes longer than that of Road 2, resulting in a negligible share of only about 2\%. Nonetheless, Road 2's share increases from 69\% to 74\%, indicating that the mere presence of Road 3, although rarely chosen, enhances the relative attractiveness of Road 2. This pattern becomes even more evident under Setting C: after Road 3 is introduced, fewer than 1\% of travelers select it, as it is more expensive and approximately five minutes slower than Road 2. Nevertheless, Road 2 becomes markedly more popular, with its share rising from 55\% to 66\%. These patterns, however, are not reproduced by the calibrated scalar-cost UE benchmark. Under this fixed generalized-cost representation, the inferior Road 3 receives no flow and does not increase Road 2's share: the predicted shares remain 31\%/69\%/0\% in Setting A and 45\%/55\%/0\% in Setting C. Thus, the LLM-guided dynamics generate a share shift that is consistent with decoy-effect logic: an inferior alternative is rarely chosen, but its presence makes the option that dominates it appear more attractive.

The response analysis in Appendix~\ref{app:tolling-response} helps explain this pattern. In brief, the representative agent often compares the two tolled routes first and recognizes that Road 2 dominates Road 3 in both time and toll. This clear semantic comparison leads Road 2 to be repeatedly reinforced, while Road 1 is evaluated mainly as the slower but toll-free alternative. This reasoning is related to the logic behind existing mathematical models that can capture decoy effects, such as context-dependent random-utility models \citep{guevara2016modeling}. Such models explicitly introduce comparison or context-dependent terms so that an alternative's attractiveness can depend on its relationship to other options in the choice set. In our framework, however, this comparison structure is not specified in advance, but emerges from the LLM agent's natural-language reasoning over the available routes.

Overall, the experiment should be interpreted as a diagnostic proof-of-concept. The LLM-guided dynamics and the calibrated scalar-cost UE benchmark produce similar qualitative predictions when the added route is genuinely competitive, but differ when the added route is inferior and comparison effects may matter. For practical use, such disagreement should not be interpreted as automatic evidence that the LLM prediction is correct; rather, it indicates that the policy scenario may involve context-dependent behavioral responses that deserve closer empirical investigation.

\begin{remark}[Baseline stability]
\label{rm:baseline-tolling}
To check whether baseline instability persists in this richer setting, we test the fully LLM-driven baseline and the fully LLM-driven RL baseline under Setting A without Road~3 (Appendix~\ref{app:baseline-tolling}). Both oscillate throughout the simulated horizon without settling into stable strategies, consistent with the classic traffic assignment results. We therefore do not extend these baseline tests to the remaining richer scenarios.
\end{remark}

\subsection{Transit transfer under denied-boarding risk}
\label{sec:risk}

The experiment is conducted on a network with a single OD pair connecting the CBD workplace to a residential area after work. The total travel demand is set to $d = 50$. All travelers are homogeneous and choose between two commuting options.

\begin{itemize}[leftmargin=*, topsep=0pt, itemsep=0pt, parsep=0pt]

    \item \textbf{Option 1: Express shuttle with mini-bus transfer.} The traveler first takes an express shuttle from the CBD to a transit center. The shuttle is available once after work and takes $t_s$ minutes. Upon arrival at the transit center, the traveler transfers to a mini-bus heading to the residential area. The mini-bus arrives every $H$ minutes, has a capacity of 25 passengers, and the ride itself takes $t_b$ minutes. In normal conditions, the shuttle arrives around the time when a mini-bus is available. However, if too many commuters choose this transfer option and the mini-bus is full upon arrival, the traveler is denied boarding and must wait for the next mini-bus.

    \item \textbf{Option 2: Direct subway.} The traveler can alternatively take a direct subway from the CBD to the residential area. The subway runs frequently, so the traveler can usually board almost immediately. However, because the route involves a detour, the total travel time is fixed at 45 minutes.

\end{itemize}

Since the total demand is $50$ and the mini-bus capacity is $25$, at most one mini-bus is missed in this stylized setting. Given $\evf_1$ (the demand choosing Option 1), the share of Option 1 demand that is denied boarding is 0 if $\evf_1 \leq 25$ and $r(\evf_1)=(\evf_1 - 25) / \evf_1$ if $\evf_1 > 25$. Let $\tau=t_s+t_b$ denote the total travel time when the traveler boards the first mini-bus. The travel-time distribution of Option 1 is then
$$
T_1(\evf_1)=
\begin{cases}
\tau, & \text{with probability } 1-r(\evf_1),\\
\tau+H, & \text{with probability } r(\evf_1).
\end{cases}
$$
We first vary $\tau$ and $H$ to examine responses to travel-time risk by widening the gap between successful-boarding and denied-boarding outcomes. We then hold the successful-boarding time fixed and compare how the LLM-guided framework and conventional risk models respond when the delayed trip changes from being slower than the subway to taking the same time.

\subsubsection{Implementation}

In this experiment, $|\mathcal{M}|=1$, and $K_m$ consists of the two available commuting options.
Accordingly, only one representative LLM agent is created. The scenario-dependent component of the system prompt describes the two options, and an additional instruction is similarly appended to the requirement block to guide the agent toward more human-like boundedly rational reasoning. The feedback prompt, however, is slightly different from the previous experiments. Since denied boarding is a stochastic event, it would be unnatural to tell the agent that the ``probability'' of denied boarding is a certain value after each day's experience has been realized.  Instead, we translate the denied-boarding probability induced by the current aggregate demand into a natural-frequency description over a recent period. For example, if the denied-boarding probability is $0.1$, the prompt describes the experience as one in which the traveler was denied boarding on $10\%$ of recent commuting days and boarded immediately on the remaining days. The exact formats of the scenario prompt, additional instruction, and feedback prompt can be found in Appendix~\ref{app:prompt-risk}.

\subsubsection{Results} 

Under this setting, we perform two sets of experiments. The first examines how travelers' preferences for the two options change as the separation between successful-boarding and denied-boarding travel times increases; the second examines how these preferences change as the waiting time after denied boarding decreases.

\begin{figure}[ht]
\centering
\includegraphics[width=1\textwidth]{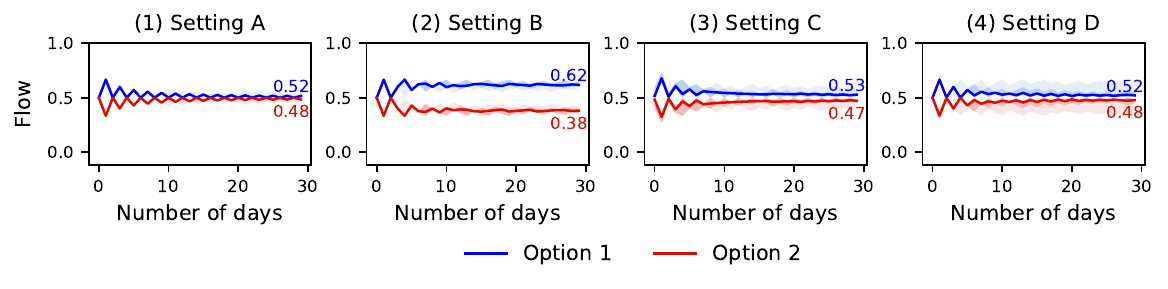}
\caption{Evolution of option shares under the LLM-guided RL dynamics for the four denied-boarding-risk settings. Solid lines denote trimmed central-tendency trajectories; darker and lighter shaded areas denote the six-run and ten-run spreads, respectively.}
\label{fig:risk-additional}
\end{figure}

In the first set of experiments, we consider four settings with $(t_s,t_b,H)$ equal to $(20,22,10)$, $(20,20,14)$, $(20,18,18)$, and $(20,16,22)$ minutes, respectively. The corresponding travel times under successful/denied boarding are 42/52, 40/54, 38/56, and 36/58 minutes. Across Settings A--D, the overall travel-time level remains broadly comparable, while the gap between the two possible bus outcomes gradually increases from 10 to 22 minutes. In all four settings, the transfer option is faster than the 45-minute subway when boarding succeeds but slower when boarding is denied. In Figure~\ref{fig:risk-additional}, we report the LLM-guided trajectories for the four settings, where the dynamics can be seen to stabilize quickly, typically within about 20 iterations. 

Here, an interesting non-monotonic pattern emerges as the successful-boarding and denied-boarding travel times become more widely separated: the final transfer-option share first rises from about 52\% in Setting~A to 62\% in Setting~B, and then falls to about 53\% and 52\% in Settings~C and~D. Evaluated at the reported final shares, the corresponding expected travel times are approximately 42.4, 42.7, 39.0, and 36.8 minutes. Setting~B is particularly noteworthy: its higher transfer share relative to Setting~A coexists with both a higher expected travel time and greater variability. This pattern contrasts with the simple intuition that greater uncertainty necessarily reduces an option's attractiveness. A possible explanation is that a modest degree of risk can be attractive when it offers a meaningful chance of a shorter trip, while this attraction weakens as the potential extra wait becomes more severe. Indeed, this interpretation has empirical grounding. A similar non-monotonic response was observed by \citet{yang2017experimental} in a repeated mode-choice experiment, where participants repeatedly chose between bus and subway while learning from uncertain travel experiences. With expected travel times held fixed across settings, the bus choice share first increased and then decreased as its travel-time variability grew. Our framework generates a qualitatively similar pattern without explicitly prescribing it, which provides proof-of-concept evidence of human-like responses to travel-time risk.

\begin{figure}[ht]
\centering
\includegraphics[width=0.7\textwidth]{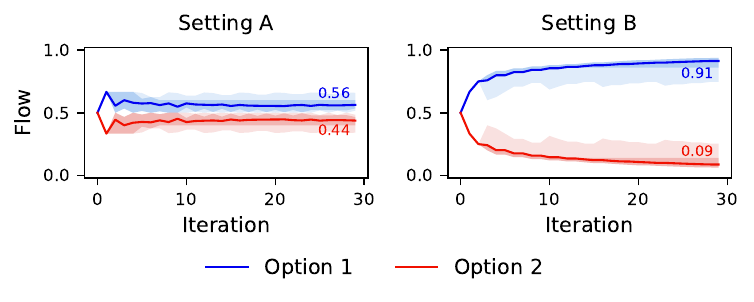}
\caption{Evolution of route flows under the LLM-guided RL dynamics for the two transit transfer settings. Solid lines denote trimmed central-tendency trajectories; darker and lighter shaded areas denote the six-run and ten-run spreads, respectively.}
\label{fig:exp-risk-results}
\end{figure}

In the second set of experiments, we compare how the LLM-guided framework and conventional risk models respond when denied boarding no longer makes the transfer option slower than the subway. We set $t_s=10$ and $t_b=20$, so successful boarding always takes $\tau=30$ minutes. Two headway values are then considered for $H$. In Setting A, the headway is $H=22$ minutes. Thus, if the mini-bus demand exceeds capacity and a traveler is denied boarding once, the realized travel time of Option 1 becomes $10+22+20=52$ minutes, which is worse than the 45-minute direct subway option. In Setting B, the headway is $H=15$ minutes. In this boundary case, one denied-boarding event makes the realized travel time of Option 1 equal to $10+15+20=45$ minutes, exactly the same as the direct subway option. These two settings allow us to compare how the LLM-guided framework and conventional risk models respond when denied boarding changes from a harmful outcome to one that is no worse than the direct alternative. In Figure~\ref{fig:exp-risk-results}, we report the LLM-guided trajectories, where it can be seen that the trajectories stabilize quickly across both settings, typically within about 10--20 iterations. As the consequence of denied boarding becomes less severe from Setting A to Setting B, the final share of the transfer option increases sharply: about 56\% in Setting A and 91\% in Setting B.

Next, we apply conventional risk-modeling approaches to the same setting. For these two targeted settings, we compare the LLM-guided dynamics with three risk-scalarized UE benchmarks. The first is a worst-case criterion, where Option 1 is evaluated by its worst realized travel time. The second is a CVaR-based criterion, with $\alpha=0.95$ as a reference confidence level, where
\[
C_1^{\mathrm{CVaR}}(\evf_1)
=30+H\cdot\min\{r(\evf_1)/(1-\alpha),1\}.
\]
The third is a mean--variance criterion, where
$
C_1^{\mathrm{MV}}(\evf_1)
=E[T_1(\evf_1)]+\gamma\cdot\mathrm{Var}[T_1(\evf_1)].
$
For each criterion, we replace the uncertain travel time of Option 1 with the corresponding scalar cost, set the cost of Option 2 to 45 minutes, and apply the UE condition. The benchmark results are as follows.
\begin{itemize}[leftmargin=*, topsep=0pt, itemsep=0pt, parsep=0pt]

    \item \textbf{Worst-case criterion.} Under the worst-case criterion, Option 1 is evaluated by its worst realized travel time. In Setting A, $H=22$, so the worst-case cost of Option 1 jumps from 30 to 52 minutes once its demand exceeds the mini-bus capacity. As a result, the discontinuity in the worst-case cost creates no stable flow split satisfying the standard UE condition: if $\evf_1 \leq 25$, Option 1 is strictly better than Option 2, so travelers using Option 2 would prefer to switch to Option 1; if $\evf_1 > 25$, Option 1 becomes worse than Option 2 under worst-case evaluation, so travelers using Option 1 would prefer to switch back to Option 2. In Setting B, $H=15$, so once capacity is exceeded, the worst-case cost of Option 1 becomes 45 minutes, exactly equal to the direct subway. Therefore, any allocation with Option 1 demand above capacity can satisfy the UE condition, yielding a wide range of equilibrium shares for Option 1, from just above 50\% up to 100\%. Thus, the worst-case criterion provides no sharp behavioral prediction.

    \item \textbf{CVaR criterion.}
    The CVaR criterion provides a less extreme tail-risk representation than the pure worst-case rule, but its prediction depends critically on the chosen confidence level $\alpha$. It can be checked that when $\alpha \approx 0.843$, the CVaR-based scalar cost of Option 1 evaluated at the LLM-guided flow in Setting A is exactly 45 minutes, equal to the cost of Option 2. Thus, with this calibrated value of $\alpha$, the CVaR benchmark reproduces the LLM-guided prediction in Setting A. However, when the same $\alpha$ is applied to Setting B, the CVaR benchmark no longer gives a sharp prediction. In Setting B, once the denied-boarding share is large enough for the CVaR cost to reach 45 minutes, both options have equal scalar cost. With $\alpha \approx 0.843$, this occurs when the Option 1 share is at least about 59\%. Hence, any Option 1 share in the range from roughly 59\% to 100\% can satisfy the UE condition. For an $\alpha$ value more commonly adopted in tail-risk modeling, such as $\alpha=0.95$, the CVaR benchmark becomes more conservative in Setting A: only $51.8\%$ choose Option 1. In Setting B, the same $\alpha$ again fails to give a sharp behavioral prediction: any Option 1 share between roughly $52.6\%$ and $100\%$ can satisfy the UE condition.

        \item \textbf{Mean--variance criterion.}
    The mean--variance criterion depends on the variance-penalty parameter $\gamma$. Compared with the confidence level $\alpha$ in the CVaR criterion, $\gamma$ is less directly interpretable, as its behavioral meaning depends on the units and scale of the travel-time variability. It can be easily checked that when $\gamma \approx 0.273$, the mean--variance cost of Option 1 is exactly equal to the 45-minute cost of Option 2. Thus, with this value of $\gamma$, the mean--variance benchmark reproduces the LLM-guided prediction in Setting A. However, when the same $\gamma$ is applied to Setting B, the mean--variance benchmark predicts a much lower transfer-option share than the LLM-guided dynamics, namely an Option 1 share of about only 66\%.
\end{itemize}
The comparison can be summarized as follows. In Setting A, the LLM-guided outcome can be reproduced by scalar risk models under suitable parameter choices: the CVaR model matches it when $\alpha \approx 0.843$, and the mean--variance model matches it when $\gamma \approx 0.273$. This suggests that, viewed through conventional scalar-risk lenses, the LLM-guided behavior in Setting A is consistent with risk aversion, but not with an extremely conservative form of risk aversion. We do not claim that the LLM-guided prediction in Setting A is ``correct." Rather, the point is that the LLM-guided outcome lies in a behaviorally plausible middle range.

The more revealing comparison arises in Setting B. In this setting, Option 1 takes 30 minutes when the first mini-bus is boarded and 45 minutes when denied boarding occurs, while Option 2 always takes 45 minutes. Thus, in terms of realized travel time, Option 1 is never worse than Option 2 and is often strictly better. The LLM-guided dynamics respond accordingly: the share of Option 1 increases sharply to about 91\%. While, Option 2 still remains only as a small reliable fallback, because some travelers found worrying to be denied boarding and then waiting for a long time to be a real ``hassle," in their own words. The mean--variance benchmark, however, is somewhat problematic here. While $\gamma \approx 0.273$ is reasonable (at least not outrageous) in Setting A, here it predicts only about 66\% on Option 1 in Setting B, even though the transfer option is never slower and is often substantially faster. Intuitively, if one wants the mean--variance model to behave more reasonably in Setting B, a smaller $\gamma$ would be needed, because the ``risk" here is less harmful now. For CVaR, the issue is that it fails to give a sharp prediction. With the calibrated value $\alpha \approx 0.843$, any Option 1 share from roughly 59\% to 100\% can satisfy the UE condition in Setting B. 

This comparison is not meant to suggest that CVaR or mean--variance models are invalid. Rather, the point is that fixed-form, fixed-parameter risk models may produce reasonable outcomes in one scenario but become less informative or behaviorally questionable in another. This distinction is straightforward for human-like reasoning: missing a mini-bus is a serious downside when it makes the trip worse than the direct subway, but it is much less concerning when the delayed trip is still no worse than the direct alternative. The LLM-guided mechanism captures this distinction through natural-language interpretation of the travel experience, without requiring the modeler to switch risk functionals or retune parameters across settings. Appendix~\ref{app:risk-response} further illustrates this reasoning through example agent responses.

Together, these experiments provide proof-of-concept evidence of the framework's potential to capture human-like risk responses and adapt across travel contexts. The first set generates a non-monotonic choice pattern qualitatively consistent with human observations. The second shows how the framework adjusts its evaluation when the consequences of denied boarding change, while the conventional benchmarks highlight limitations of transferring fixed risk specifications across settings. Both patterns arise through the same LLM-guided learning mechanism, without explicitly prescribing these responses or retuning risk parameters for each setting.

\begin{remark}[Robustness to the LLM core]
\label{rm:llm-core}
The two transit experiments above illustrate human-like responses to travel-time uncertainty. One may naturally ask whether these findings are robust to the LLM core. We therefore repeat both experiments, keeping all other parameters unchanged, with another commercial LLM, Qwen-3.8-Max \citep{qwen38max}. We find that both qualitative patterns are preserved, which provides evidence of robustness across the two tested LLM cores (Appendix~\ref{app:llm-core}).
\end{remark}

\begin{remark}[Comparison with human experimental data]
\label{rem:human-risk-comparison}
To provide a more direct comparison with human behavior, we also apply the framework using the travel-time and fare specifications of \citet{yang2017experimental}. Across four settings, the bus and subway have fixed mean travel times of 45 and 40 minutes and fares of 3 and 5 RMB, respectively. The subway travel-time standard deviation remains at 1.5 minutes, while the bus standard deviation increases from 1 to 2, 5, and 7.5 minutes. The framework generates final bus shares of approximately 43\%, 62\%, 44\%, and 41\%, compared with the reported human choice proportions of 43.9\%, 50.7\%, 44.3\%, and 34.2\%. Without calibration to these observations, the framework reproduces the tendency for bus choice to first increase and then decrease as travel-time variability grows. This provides further evidence of qualitative agreement with human behavior, although quantitative differences remain.
\end{remark}

\subsection{Multi-modal commuting}
\label{sec:multi-modal}

\begin{figure}[ht]
    \centering
    \includegraphics[width=0.9\textwidth]{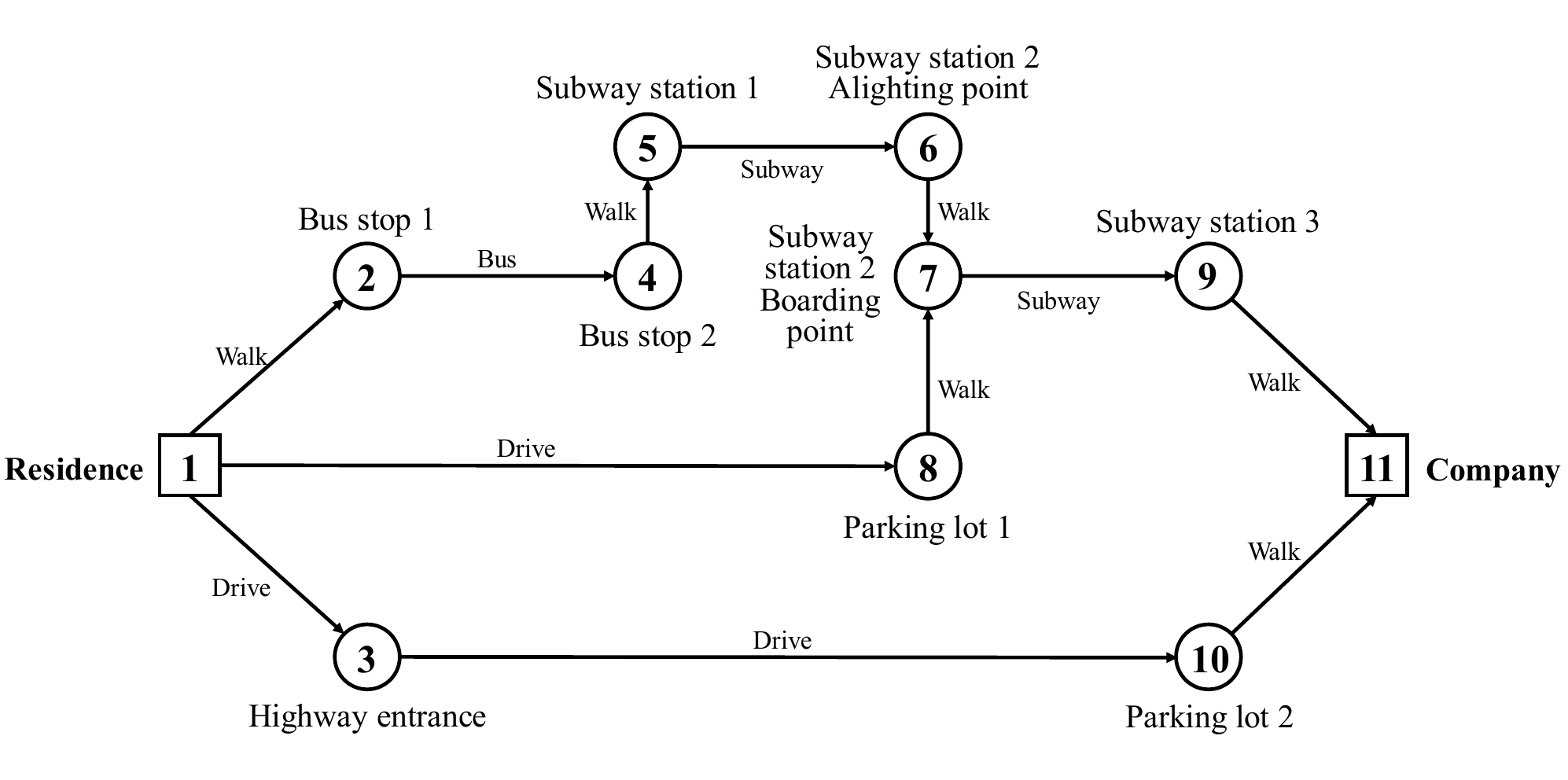}
    \captionof{figure}{An illustration of the three commuting mode choices.}
    \label{fig: mode choice}
\end{figure}

The experiment is conducted on a stylized network with a single OD pair that connects a suburban residential area to a downtown workplace (see Figure~\ref{fig: mode choice}). To represent travelers with different socioeconomic backgrounds, we divide the demand into three stylized profiles: a high-income group ($d_1=7.5$, monthly salary \(\$12000\)), a middle-income group ($d_2=15$, monthly salary \(\$7000\)), and a low-income group ($d_3=7.5$, monthly salary \(\$2000\)). The three profiles differ not only in income but also in the broader socioeconomic context. High-income travelers are assumed to own higher-value housing and vehicles and to face larger monthly mortgage payments; middle-income travelers own a more modest vehicle and condo with lower mortgage payments; and low-income travelers own a lower-value vehicle and rent an apartment.

All travelers face the same set of commuting options, which correspond to three different modes.
\begin{itemize}[leftmargin=*, topsep=0pt, itemsep=0pt, parsep=0pt]
    \item \textbf{Public transit.} The commuter first walks 10 minutes along an exposed route to a bus stop (Node 2), waits about 3 minutes, and rides for 10 minutes to an intermediate stop (Node 4) near a metro station (Node 5). The transfer to the metro takes approximately 5 minutes, including walking and waiting. The metro journey then consists of 13 minutes on the first line (Node 5 $\to$ Node 6) and 10 minutes on the second line (Node 7 $\to$ Node 9), connected by a 4-minute transfer between lines. Finally, walking from the terminal station to the workplace (Node 9 $\to$ Node 11) takes an additional 3 minutes. The total one-way fare is \$8 (\$3 for the bus and \$5 for the metro). Crowding levels on both bus and metro segments increase with passenger volume.
    \item \textbf{Driving.} The commuter first drives along local city roads (Node 1 $\to$ Node 3) before entering the highway segment (Node 3 $\to$ Node 10). The estimated fuel cost for the one-way trip is \$4. Upon arrival, the vehicle is parked in a downtown parking lot (Node 10), where the daily fee is \$40. Finally, the commuter walks about 3 minutes from the parking lot to the workplace.
    \item \textbf{Park-and-ride.} The commuter first drives to a parking lot (Node 8), which is adjacent to the metro station offering a direct line to downtown (Node 7). Since this area lies outside the city center, traffic congestion is minimal, and the parking fee is lower than downtown (\$25 per day). The fuel cost for this segment is \$2. After parking, the commuter spends about 3 minutes walking to the metro station and waiting for the next train. The remaining journey follows the same metro route as in the public transit option.
\end{itemize}
The travel times on all road segments follow the BPR form
$u_{\text{time}, a}(\evx_a) = \evu_{0, a} \cdot (1 + 0.15 \cdot (\evx_a / \evv_{a})^{4})$, where $a = (1, 3), (3, 10), (1, 8).$ We set $\evu_{0,(1,3)} = 5$ and $\evv_{(1,3)} = 8$; $\evu_{0,(3,10)} = 25$ and $\evv_{(3,10)} = 10$; $\evu_{0,(1,8)} = 15$ and $\evv_{(1,8)} = 12$.
For each transit line, a fixed number of \textit{empty} seats at departure is given. For the bus and the trains of the two metro lines, these capacities are assumed to be 5, 3.75, and 7.5 units of commuters, respectively. All other commuters may still board the vehicle, but must stand. Once a commuting mode is chosen for the trip to work, the commuter has to use the same mode to return. The experience on the way home is assumed to be identical to that of the outbound journey.

This scenario is a stylized proof-of-concept rather than a calibrated representation of a specific city or a real commuter population. Here, our aim is to test whether income-dependent mode-choice patterns can emerge without imposing a fixed generalized-cost formula or income-specific utility weights. The representative agents update their strategies independently and do not observe the decisions of other income groups. A key question is whether the resulting dynamics can still generate interpretable mode shares that align with the socioeconomic profiles of the represented travelers.

\subsubsection{Implementation}
\label{sec:multi-modal-results}

As the network contains a single OD pair and commuters are divided into three income-based groups, $|\gM| = 3$, and for each class $m \in \gM$, the set $\gK_m$ corresponds to the same three available commuting modes. The scenario-dependent prompt is customized by income class. In addition to monthly salary, each representative agent is given decision-relevant socioeconomic background information, such as car ownership, vehicle value, housing type, and housing costs. The requirement block also includes additional instructions, such as instructing the agent to behave as a boundedly rational human, similar to the previous experiments.  The daily feedback $e_{m,k}(\vf^t)$ for each option reports information for every segment of the trip separately. It covers multiple dimensions, including travel time, monetary expenses, transfer experiences, and how many are seated and how many are standing. The exact formats of the scenario prompt, additional instruction, and feedback prompt are provided in Appendix~\ref{app:prompt-multi-modal}.

Because the multi-modal commuting scenario involves substantially richer and more profile-dependent reasoning than the previous experiments, the main results use the voting-based qualitative reasoning procedure described in Remark~\ref{rm:voting}. Specifically, for each representative agent on each simulated day, we query the LLM 10 times under the same dialog history and feedback prompt, and determine the final positively reinforced set by majority voting. The subsequent mixed-strategy update remains unchanged and follows Rule~1. Taking this voting-based implementation as the main specification, we further conduct three ablation checks: (1) we remove the voting step and use a single LLM response to examine whether voting reduces run-to-run variation; (2) we apply the bounded-memory design discussed in Remark~\ref{rm:memory} to test whether retaining only recent dialog history materially affects the resulting dynamics; (3) we remove detailed socioeconomic background information, such as the values of their vehicles, and provide only income levels, to examine whether richer profile information improves the stability and reduce run-by-run variance.

\subsubsection{Results}

\begin{figure}[ht]
    \centering
    \vspace{-5pt}
    \includegraphics[width=0.95\textwidth]{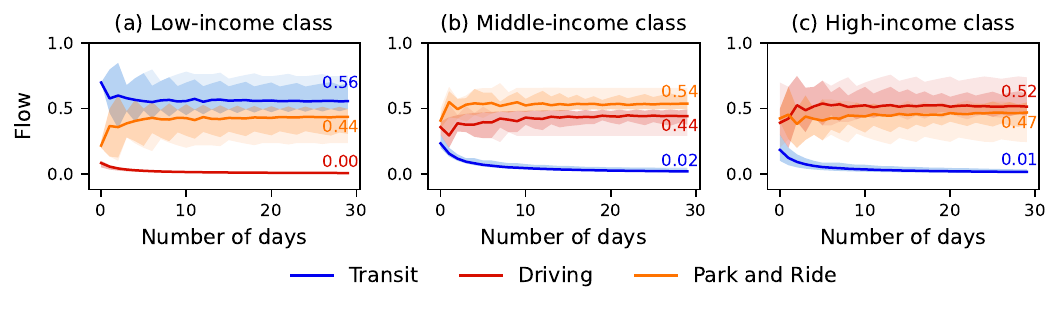}
    \captionof{figure}{Evolution of route flows over simulated days under the LLM-guided RL dynamics with voting-based qualitative reasoning for the multi-modal commuting scenario. { Solid lines denote trimmed central-tendency trajectories; darker and lighter shaded areas denote the six-run and ten-run spreads, respectively.}}
    \label{fig:exp3-result}
\end{figure}

Figure~\ref{fig:exp3-result} illustrates the evolution of the mode shares under the LLM-guided RL dynamics. 
{ The run-to-run variation is modest under the voting-based implementation: the trimmed trajectories exhibit very small dispersion, especially for the middle- and high-income classes, while the full ten-run spread is only moderately larger. This suggests that the voting scheme helps stabilize the aggregate mode-choice dynamics while still allowing the full range of LLM-induced variation to be observed. The trajectories stabilize quickly for all three income classes, and the final patterns are economically interpretable. For the low-income class, public transit and park-and-ride dominate, with final shares of approximately 56\% and 44\%, respectively, while direct driving is negligible. For the middle-income class, park-and-ride becomes the most attractive option, with a final share of about 54\%, followed by driving at about 44\%, while transit falls to about 2\%. For the high-income class, direct driving and park-and-ride dominate, with final shares of about 52\% and 47\%, respectively, while transit is rarely selected. The most popular modes for the low-, middle-, and high-income classes are public transit, park-and-ride, and driving, respectively, reflecting an increasing willingness to trade monetary cost for comfort and convenience as income rises.

Meanwhile, park-and-ride is a popular choice for all classes, which is consistent with prior evidence that park-and-ride can attract both low- and high-income travelers \citep{zhao2019does, lachapelle2023breaking}. In the experiment, it is attractive across income groups for different reasons. For the middle-income class, it is mainly interpreted as a compromise: it is much cheaper than driving while remaining close to driving in travel time. For the high-income class, park-and-ride functions more as a hedge against bad driving days caused by congestion. For the low-income class, park-and-ride is sometimes chosen to be reinforced because it provides relief from the accumulated fatigue of full public transit. We refer the readers to Appendix~\ref{app:multi-modal-response} for a detailed discussion. 
Together with recent work on LLM-based travel preferences \citep{yan2026valuing,voltesdorta2025airline}, this result supports the potential of LLMs to represent how travelers with different profiles weigh the same travel attributes. Our experiment also adds evidence that such profile-dependent evaluations can translate into stable aggregate mode shares through day-to-day learning.

The results corresponding to the three ablation checks are reported in Appendix~\ref{app:robustness}. The main findings are summarized as follows. First, removing the voting step largely preserves the qualitative mode-choice pattern, but increases run-to-run variation, which supports the role of repeated querying and voting as a variance-control device: without voting, idiosyncratic single-day reinforcement decisions can be amplified through subsequent strategy updates and dialog history. Second, using bounded memory broadly preserves the aggregate patterns, but produces more outlying trajectories. Although bounded memory substantially reduces token usage, it does not reduce runtime in our implementation; thus, it is useful for reducing API token cost, but is not necessarily preferable when the objective is stable repeated simulation. Third, removing rich socioeconomic background information makes the low-income pattern less stable, suggesting that detailed profile anchors help the representative agent interpret their travel experience. The third finding also connects our experiment with recent persona-based LLM travel behavior studies \citep{liu2026persona,tzachristas2026pdm}, which use specified traveler profiles to guide LLMs in simulating individual travel choices. Our results further show that richer profile information can also improve the consistency of mode-choice outcomes generated by LLM simulations across repeated runs. In our experiment, these profiles are predefined rather than learned or calibrated from empirical data. Extending the framework to a specific commuter population would therefore require empirically grounded profiles and behavioral calibration.

\section{Conclusion}
\label{sec:conclusion}

This study proposed an LLM-guided, representative-agent framework for modeling day-to-day traffic dynamics, which integrates the reasoning capability of contemporary LLMs with a conventional reinforcement learning (RL) framework, rather than leaving the full behavioral update to the LLM.

Compared with prior LLM–based agent-based models (ABMs), which work by assigning a sufficient number of atomic LLM agents to each class, the proposed framework has the following advantages.
First, the representative–agent design reduces daily LLM calls from many per class in LLM-driven ABMs to just one per class, which substantially lowers both token usage and runtime cost.  Second, our framework is \textit{not} fully black-box. The LLM provides only the \emph{direction of learning}, along with a short rationale that can be audited, while the quantitative update follows an explicit, transparent rule. By contrast, prior LLM agents rely on fully LLM-driven updates.  Third, earlier LLM–based ABMs face two sources of instability: (i) unconstrained black-box updates and (ii) sampling noise from finitely many atomic agents. Our design addresses both: aggregate flows are computed directly from the representative's mixed strategy rather than Monte Carlo sampling, and step sizes are explicitly controlled by decaying the step size. Taken together, these innovations make our framework more interpretable, stable, and computationally efficient, potentially as a stronger candidate for reliable traffic modeling and policy evaluation.

The numerical results demonstrate that the proposed approach can reproduce classical equilibrium patterns in simple settings, while simultaneously capturing richer, human-like decision processes that are difficult to represent through a single pre-specified generalized-cost function in more complex scenarios. Specifically, in the classic traffic assignment setting, the LLM-guided RL mechanism approaches near-UE states; by contrast, the fully LLM-driven baselines remain unstable and do not reliably approach UE. In richer settings involving tolling, denied-boarding risk, and multi-modal commuting, the framework generates stable and interpretable patterns, including decoy-like route-choice shifts, consequence-sensitive risk responses, and income-dependent mode-choice trade-offs. These findings suggest that the proposed framework has the potential to offer a valid way to study travel decisions that are difficult to represent through a single pre-specified generalized-cost function. Although we do not claim that LLM predictions are necessarily correct in practical policy evaluation, we view the LLM as a complement to conventional models. When substantial differences arise between their predictions, this may alert planners to behavioral comparison effects or other contextual responses that deserve further empirical investigation. Once such differences are identified, observed behavioral data can be used to calibrate or fine-tune the LLM so that its predictions better reflect traveler responses.

The experiments in this paper should be viewed as stylized proof-of-concept studies rather than calibrated real-world validations. A full empirical implementation would require aligning the representative LLM agents with behavioral evidence from the target setting, just as conventional traffic and choice models require calibration of parameters such as value of time, transfer penalties, and alternative-specific constants. Such evidence may include observed route or mode choices, navigation records, stated-preference surveys, questionnaire responses about how travelers trade off time, monetary cost, reliability, comfort, transfers, and crowding, or qualitative information about traveler perceptions. For example, in a CBD commuting application similar to our experiments, one could survey commuters with known socioeconomic profiles and ask them to evaluate concrete choice situations involving a faster tolled route versus a slower toll-free route, or a usually faster transfer option with denied-boarding risk versus a reliable direct subway. The respondents' choices and stated reasoning could then be used as few-shot examples for representative agents with similar profiles. If revealed route or mode choices are available, they could also be used to fine-tune or otherwise calibrate the underlying LLM so that the model better reproduces observed baseline behavior. The aligned model could then be evaluated on held-out observations or observed responses to policy changes not used during alignment before being used for counterfactual policy analysis. We therefore regard real-world behavioral alignment and out-of-sample validation as an important next step, while the present paper focuses on introducing the methodological architecture and examining its behavior in controlled settings.

Alongside behavioral alignment, practical deployment of the proposed approach also requires accounting for stochastic variation in the simulated outcomes. In our experiments, individual runs typically stabilize, although their final shares may differ because of stochastic variation in the LLM's reinforcement decisions. Repeated simulations therefore characterize the variability of these outcomes, while averaging summarizes their central tendency. For policy benchmarking, we recommend reporting a central-tendency summary, such as a trimmed mean, together with the range across all runs, and examining whether the policy conclusions remain consistent across these outcomes.

A further methodological limitation is that the current mechanism only identifies which options should be reinforced, without distinguishing how strongly different options should be reinforced. This simple interface keeps the LLM evaluation and numerical update clearly separated, but uses only a limited amount of the information that the LLM could provide. Future work could therefore introduce graded reinforcement signals that reflect both which options are favored and how strongly. This extension could still keep the numerical strategy update governed by an explicit rule.

\bibliographystyle{apalike} 
{
\bibliography{sample} 
}

\newpage
\appendix

\section{Requirement Prompts for the Algorithms}
\label{app:additional}

This appendix reports the algorithm-specific requirement component $\omega_m^{\text{s-requirement}}$ of the system-role prompt for Algorithms~\ref{alg:baseline}, \ref{alg:baseline2}, and~\ref{alg:rl}. These requirement blocks specify the daily task, reasoning guidelines, and parsable output format used by each learning mechanism. They are not tied to any particular traffic scenario.

\vspace{10pt}
\begin{system}[$\omega_m^{\text{s-requirement}}$ when $\pi_m$ is defined according to Algorithm~2]
    \textit{Each day, you will be asked to revise your strategy by reflecting on your current strategy and your travel experiences in recent days. Your answer must follow the following requirements:
\begin{itemize}[leftmargin=*, topsep=0pt, itemsep=0pt, parsep=0pt]
  \item[-] Think step by step, starting by reflecting your current strategy and the recent travel experiences. 
  \item[-] Provide a thorough but concise analysis, and present the final result only after that.
  \item[-] Output: \result{} Updated strategy: [xx, xx, \dots]. \closeresult{}.
\end{itemize}
}
\end{system}

\begin{system}[$\omega_m^{\text{s-requirement}}$ when $\pi_m$ is defined according to Algorithm~3]
    \textit{Each day, you will complete two tasks based on your current strategy and recent travel experiences:
\begin{itemize}[leftmargin=*, topsep=0pt, itemsep=0pt, parsep=0pt]
  \item[-] Task 1: Select the commuting options you would like to use more often.
  \item[-] Task 2: Indicate how you will update your mixed strategy to reinforce the use of these options.
\end{itemize}
For Task 1, your answer must follow these requirements:
\begin{itemize}[leftmargin=*, topsep=0pt, itemsep=0pt, parsep=0pt]
  \item[-] Think step by step, starting by reflecting on your current strategy and the recent travel experiences. 
  \item[-] Provide a thorough but concise analysis, and present the final result only after that.
  \item[-] You must not increase the probability of all options simultaneously.
  \item[-] If you decide not to reinforce any options, output: \result{} None. \closeresult{}.
  \item[-] Otherwise, output: \result{} Options selected for increase: [xx, xx, \dots]. \closeresult{}.
\end{itemize}
For Task 2, your answer must follow these requirements:
\begin{itemize}[leftmargin=*, topsep=0pt, itemsep=0pt, parsep=0pt]
  \item[-] Think step by step. 
  \item[-] Provide a thorough but concise analysis, and present the final result only after that.
  \item[-] Output: \result{} Updated strategy: [xx, xx, \dots]. \closeresult{}.
\end{itemize}
}
\end{system}

\begin{system}[$\omega_m^{\text{s-requirement}}$ when $\pi_m$ is defined according to Algorithm~4]
    \textit{Each day, you will complete two tasks based on your current strategy and recent travel experiences:
\begin{itemize}[leftmargin=*, topsep=0pt, itemsep=0pt, parsep=0pt]
  \item[-] Task 1: ``Select the commuting options you would like to use more often."
  \item[-] Task 2: ``Indicate how you will update your mixed strategy to reinforce the use of these options."
\end{itemize}
For Task 1, your answer must follow these requirements:
\begin{itemize}[leftmargin=*, topsep=0pt, itemsep=0pt, parsep=0pt]
  \item[-] Think step by step, starting by reflecting on your current strategy and the recent travel experiences. 
  \item[-] Provide a thorough but concise analysis, and present the final result only after that.
  \item[-] You must not increase the probability of all options simultaneously.
  \item[-] If you decide not to reinforce any options, output: \result{} Options selected for increase: None. \closeresult{}; otherwise, output: \result{} Options selected for increase: [xx, xx, \dots]. \closeresult{}.
\end{itemize}
}
\end{system}

\section{Prompt Design for the Experiments}
\label{app:prompt-design}
This appendix reports the prompt components used to instantiate the representative LLM agents in the numerical experiments. For each experiment, we provide three elements: the system-role scenario prompt, any scenario-specific requirement items appended to the requirement block in Appendix~\ref{app:additional}, and the user-role feedback prompt. In all prompt templates, quantities enclosed in braces are placeholders that are replaced by the corresponding simulated values.

\subsection{Classic traffic assignment}
\label{app:prompt-classic}

The scenario prompt specifies the available routes and the fact that route travel times depend on congestion; the feedback prompt only reports the realized route travel-time vector for the current day.

\begin{system}[$\omega_m^{\text{s-scenario}}$ for classic traffic assignment]
    \textit{You are a daily commuter in a transportation network. Each day, you may choose from \{$|\gK_m|$\} available routes (indexed by $1, \ldots, \{|\gK_m|\}$) as your commuting option. The travel time on each route increases as more commuters use it. However, you do not know how others explore the different routes.} 
\end{system}

\begin{user}[$\omega_m^{t, \text{u-feedback}}$ for classic traffic assignment]
    \textit{Today, the travel times of the $\{|\gK_m|\}$ routes are realized as follows: $\{[c_{m, 1}(\vf^t), \ldots, c_{m, |\gK_m|}(\vf^t)]\}$.}
\end{user}

\subsection{Highway tolling}
\label{app:prompt-tolling}

The scenario prompt provides a simple socioeconomic profile so that the toll amounts can be interpreted relative to the commuter's budget, and then describes the toll status of each route. A bounded-rationality add-on requirement item is appended to guide the agent to behave like a human commuter. The feedback prompt has two versions, depending on whether Road 3 is included. In both versions, route-specific travel times and tolls are reported directly.

\begin{system}[$\omega_m^{\text{s-scenario}}$ for the highway tolling scenario]
    \textit{You are a daily commuter in a transportation network. You have a monthly salary of 25,000 HKD (the average income level in your area is about 20,000 HKD). You rent an apartment with a monthly rent of 9,000 HKD, and your monthly spending on food is approximately 5,000 HKD. Each morning, you may choose from these 3 available routes (indexed by 1, \ldots, 3) as your commuting option:
    \begin{itemize}[leftmargin=*, topsep=0pt, itemsep=0pt, parsep=0pt]
        \item[-] Route 1 is toll-free.
        \item[-] Route 2 is a tolled route costing 30 HKD per trip. 
        \item[-] {\color{gray} if Road 3 exists:} Route 3 is also tolled at 34 HKD per trip. 
    \end{itemize}
    The travel time on each route increases as more commuters use it. However, you do not know how others explore the different routes.}

\end{system}

\begin{system}[Additional items in $\omega_m^{\text{s-requirement}}$ for the highway tolling scenario]
\begin{itemize}[leftmargin=*, topsep=0pt, itemsep=0pt, parsep=0pt]
    \item[-] \textit{You should behave like a human being with bounded rationality: make decisions through a combination of subjective perception, psychological intuition, and a moderate degree of rational analysis, rather than purely logical computation or strict comparisons.}
\end{itemize}
\end{system}

\begin{user}[$\omega_m^{t, \text{u-feedback}}$ for the tolling scenario]
    \textit{{\color{gray} if Road 3 does not exist:} Today, Route 1 has a travel time of $\{c_{m, 1}(\vf^t)\}$ minutes and is toll-free and Route 2 takes $\{c_{m, 2}(\vf^t)\}$ minutes with a toll fee of 30 HKD.}

    \small
    \textit{{\color{gray} if Road 3 exists:} Today, Route 1 has a travel time of $\{c_{m, 1}(\vf^t)\}$ minutes and is toll-free, Route 2 takes $\{c_{m, 2}(\vf^t)\}$ minutes with a toll fee of 30 HKD, and  Route 3 takes $\{c_{m, 3}(\vf^t)\}$ minutes with a toll fee of 34 HKD.}
\end{user}

\subsection{Transit transfer under denied-boarding risk}
\label{app:prompt-risk}
{ 
The transit-transfer experiment presents denied boarding as part of a recurring weekly travel experience. The scenario prompt contrasts a transfer option that may be faster but capacity-constrained with a direct subway option that is slower but capacity-unconstrained. A bounded-rationality add-on requirement item is appended to make the agent interpret this risk from the standpoint of a recurring commuter. The feedback prompt converts the denied-boarding probability induced by the current demand into the share of workdays with each realized outcome.
}

\begin{system}[$\omega_m^{\text{s-scenario}}$ for the transit transfer under denied-boarding risk]
    \textit{You are a daily commuter traveling from the CBD back to your residential area after work. Each day, you may choose one of two commuting options:
    \begin{itemize}[leftmargin=*, topsep=0pt, itemsep=0pt, parsep=0pt]
        \item[-] Option 1: You first take an express shuttle from the CBD to a transit center. The shuttle is available only once after work and takes $\{t_s\}$ minutes. You then transfer to a mini-bus to your residential area. The mini-bus arrives every $\{H\}$ minutes, has a capacity for 25 passengers, and the ride takes $\{t_b\}$ minutes. In normal conditions, the shuttle arrives at the bus stop just around the time when a mini-bus to your residential area is available. However, if too many commuters choose this option and the mini-bus is full when you arrive, you may be denied boarding and must wait for the next mini-bus. In general, even if you miss the first bus, you will be able to board the second one.
        \item[-] Option 2: You take a direct subway from the CBD to your residential area. The subway runs frequently, so you can usually board almost immediately. Because the route involves a detour, the total trip takes 45 minutes.
    \end{itemize}
    The denied-boarding risk for Option 1 generally increases when more commuters choose it. However, you do not observe how other commuters explore these two options.}
\end{system}

\begin{system}[Additional items in $\omega_m^{\text{s-requirement}}$ for the transit transfer under denied-boarding risk]
\begin{itemize}[leftmargin=*, topsep=0pt, itemsep=0pt, parsep=0pt]
    \item[-] \textit{You should behave like a human being with bounded rationality: make decisions through a combination of subjective perception, psychological intuition, and a moderate degree of rational analysis, rather than purely logical computation or strict comparisons.}
    \item[-] \textit{Your decision should take into account the likely consequences of each option.} 
\end{itemize}
\end{system}

\begin{user}[$\omega_m^{t, \text{u-feedback}}$ for the transit transfer under denied-boarding risk]
    \textit{This week, the traveler's experiences of choosing the two commuting options are realized as follows.}

    \small
    \begin{itemize}
        \item[-] \textit{For option 1 (express shuttle + mini-bus): You first took the company shuttle from your workplace to the bus stop, which took $\{t_s\}$ minutes. On $\{1-r(q)\}$ of the days, after arriving at the bus stop, you boarded the first bus without additional waiting at the bus stop; the bus ride to the residential area took $\{t_b\}$ minutes, and your total travel time was $\{t_s+t_b\}$ minutes. On $\{r(q)\}$ of the days, after arriving at the bus stop, the first bus was full, so you could not board it; you waited $\{H\}$ minutes for the next bus, then took a $\{t_b\}$-minute bus ride to the residential area, and your total travel time was $\{t_s+t_b+H\}$ minutes.
        \item[-] For option 2 (direct subway): You took the direct subway from the CBD to the residential area. The total travel time was always 45 minutes.}
    \end{itemize}
\end{user}

\subsection{Multi-modal commuting}
\label{app:prompt-multi-modal}

The multi-modal commuting experiment varies the socioeconomic profile in the scenario prompt. The add-on requirements similarly instruct the agent to reason as a boundedly rational commuter over multiple travel attributes. The feedback prompt reports each trip segment separately, including waiting, transfers, crowding, travel time, and fares, rather than aggregating the experience into a single generalized cost.
\begin{system}[$\omega_m^{\text{s-scenario}}$ for the multi-modal commuting scenario]
\textit{You are a daily commuter living in a suburban area near Chicago, and you need to travel to Downtown Chicago for work.}
\begin{itemize}[leftmargin=*, topsep=0pt, itemsep=0pt, parsep=0pt]
    \item \textit{{\color{gray}  If $m = 1$ (high-income class):} You own a private car valued at \$30,000 and a house valued at \$300,000, with a monthly mortgage payment of about \$2,500. Your monthly salary is \$12,000, higher than the average income in your area (about \$7,000).}  
    \item \textit{{\color{gray} If $m = 2$ (middle-income class):}  You own a private car valued at \$20,000 and a condo valued at \$100,000, with a monthly mortgage payment of about \$2,000. Your monthly salary is \$7,000, on par with the average income level in your area (about \$7,000).}
    \item \textit{{\color{gray} If $m = 3$ (low-income class):} You own a private car valued at \$8,000 and rent an apartment with a monthly rent of \$1,000. Your monthly salary is \$2,000, lower than the average income in your area (about \$7,000).}
\end{itemize}
\textit{Each working day, you can commute between home and work using one of the following three options.}
\begin{itemize}[leftmargin=*, topsep=1pt, itemsep=1pt, parsep=1pt]
    \item[-] \textit{Option 1 (public transit). There is no direct transit connection between your home and workplace. Your commute begins with an uncovered 0.5-mile walk to the bus station, which can be uncomfortable in extreme weather. After the bus ride to the metro station, you must transfer to another metro line to complete your journey. The ticket prices for bus and transit are \$3 and \$5 per trip, respectively.}   
    \item[-] \textit{Option 2 (driving). You can drive directly between home and work. The route to work begins with local city roads, which are slower but less congested, followed by a highway segment that can become crowded near downtown during the peak. There is an underground parking lot at your workplace (\$40 each day).}  
    \item[-] \textit{Option 3 (park-and-ride). You can also drive to a metro station located outside the downtown area, where congestion is light. A parking lot near the station is available at a much lower fee, \$25 per day, compared with downtown parking. From there, you can then take a direct metro line (without transfer) to work (\$3 per trip).}
\end{itemize}

\end{system}

\begin{system}[Additional items in $\omega_m^{\text{s-requirement}}$ for the multi-mode scenario]
\begin{itemize}[leftmargin=*, topsep=0pt, itemsep=0pt, parsep=0pt]
    \item[-] \textit{You should behave like a human being with bounded rationality: make decisions through a combination of subjective perception, psychological intuition, and a moderate degree of rational analysis, rather than purely logical computation or strict comparisons.}
    \item[-] \textit{Your decision should be based on your personal preferences and situations, considering multiple factors, including but not limited to travel time, monetary cost, and comfort for each mode.}
\end{itemize}
\end{system}

\begin{user}[$\omega_m^{t, \text{u-feedback}}$ for the multi-modal commuting scenario]
\textit{Today, the traveler's experiences of choosing the three commuting options are realized as follows.}
\begin{itemize}[leftmargin=*, topsep=1pt, itemsep=1pt, parsep=1pt]
    \item[-] \textit{Option 1 (transit). The commute to work begins with an exposed 10-minute walk (0.5 miles) to the bus stop, followed by a 3-minute wait and a 10-minute bus ride, with \{\textit{Description of crowding status}\}. The transfer to the metro takes about 5 minutes in total, including walking and waiting for the next train. The metro journey consists of 13 minutes on the first line and another 10 minutes on the second line, with a 4-minute transfer between them. The first train has \{\textit{Description of crowding status}\}, while the second has \{\textit{Description of crowding status}\}. Finally, there is a 3-minute walk from the final station to the workplace. The total monetary cost for the trip is \$8. The return trip home is similar in experience.}
    \item[-] \textit{Option 2 (driving). Driving to work takes approximately \{$u_{\text{time},(1,3)}(\vf^t) + u_{\text{time},(3,10)}(\vf^t)$\} minutes in total, including \{$u_{\text{time},(3,10)}(\vf^t)$\} minutes on the highway, where the ratio of experienced travel time to free-flow time is \{$u_{\text{time},(3,10)}(\vf^t) / \evu_{0,(3,10)}$\}. Upon arrival, a 3-minute walk from the parking lot to the workplace is required. Parking for the entire day costs \$40. The return driving time is similar, and the total fuel cost for the day is about \$8.}
    \item[-] \textit{Option 3 (park-and-ride). The drive from home to the metro station takes \{$u_{\text{time}, (1,8)}(\vf^t)$\} minutes, with a parking fee of \$25 for the day. It then takes about 3 minutes to walk to the station and wait for the next train. The metro ride to the workplace takes 10 minutes and has \{\textit{Description of crowding status on public transport}\}. Upon arrival, walking from the station to the workplace takes 3 minutes. The return trip home involves comparable driving and transit times.  The total fuel cost for the day is about \$4, and the round-trip transit fare is \$6.}
\end{itemize}
   
\end{user}

\medskip
\begin{remark}[Rationale behind the bounded-rationality instruction]
    We end this appendix with a remark on the bounded-rationality instruction used in the richer application experiments. In game-theoretic modeling, it is useful to distinguish between perfect rationality, which assumes that decision makers can precisely evaluate all relevant information and optimize accordingly, and bounded rationality, which recognizes that human travelers may rely on subjective perception, limited attention, approximate comparisons, tolerance bands, habits, and other decision heuristics \citep{simon1955behavioral}. This distinction matters for LLM-based agents. If no rationality mode is specified, the agent's decision logic may become ambiguous. In our pilot trials, we observed that an LLM may sometimes behave like an idealized analyst rather than a human traveler. For example, in the highway tolling scenario, the agent may infer a value of time, compute a generalized cost for each route, and then reason as if it were a perfectly rational cost minimizer. For this reason, in the tolling, transit-risk, and multimodal experiments, we explicitly instruct the representative agents to behave as boundedly rational human travelers. We further specify that their decisions should combine subjective perception, psychological intuition, and a moderate degree of rational analysis, which is intended to make the bounded-rationality instruction more robust: without it, the agent may still default to a formulaic generalized-cost calculation.
\end{remark}

\section{Representative Excerpts from LLM Agent Responses}
\label{app:response}
This appendix provides representative excerpts from LLM agents' responses generated under the LLM-guided RL mechanism, which complement the response analysis presented in the main body.

\subsection{Classic traffic assignment}
\label{app:classic-response}

We first present three examples of the agent's response in the classic traffic assignment scenario under the LLM-guided RL mechanism, which demonstrates that the LLM's learning is more complex than the simple best-response rule.

The first response reveals two features. First, the decision is not purely reactive to the most recent feedback; instead, the agent explicitly recalls earlier experiences and bases its decisions on consistent past performance. Second, it does not simply reinforce the single best-performing route but also maintains moderate exploration by favoring multiple promising options. These traits differ from best-response dynamics, which respond only to the previous day's costs and always reinforce the current best route. 

\begin{agent}[$\omega_m^{t, \text{a-positive}}$ (the first example in the classic traffic assignment scenario)]
\textit{My current strategy is [0.191, 0.287, 0.408, 0.114].  
Today's costs:  
Route 1: 3880.1 (worse than yesterday's 3689.7).   
Route 2: 3679.7 (second best).   
Route 3: 3657.7 (best today).   
Route 4: 3902.1.}

\smallskip
\textit{Comparing with previous days: Route 3 is again the best, {improving from 3701.0 yesterday}. Route 2 is second best, {also improved from 3785.3 yesterday}. {Routes 1 and 4 performed poorly today after yesterday's improvements, showing volatility}. Route 3 showed the best performance today and improved, so it deserves reinforcement. Route 2 also improved significantly, so I'll reinforce it as well to keep exploring both top options.}
\end{agent}

The second response illustrates an exploration-preserving case. Although Route 1 has a clear and consistent cost advantage, the agent chooses not to reinforce it further because its probability is already high and Route 2 would otherwise become too unlikely to be explored. This differs from a one-day best-response rule, which would continue reinforcing the current lowest-cost route. The response shows that the LLM can balance exploitation of a reliable route with the need to preserve minimal exploration in the mixed strategy.
\begin{agent}[$\omega_m^{t, \text{a-positive}}$ (the second example in the classic traffic assignment scenario)]
\textit{My current strategy is [0.871, 0.129].  
Today’s costs: Route 1 = 25.7, Route 2 = 31.0.  
Route 1 again performed better than Route 2, with a gap of 5.3 min--moderate but consistent. From recent days: Route 1 has been better on most days. The gap remains moderate, but Route 1 is clearly more reliable. Route 1 continues to be the preferred option, but Route 2’s probability is already low (0.129). Further increasing Route 1’s probability would push Route 2’s probability below 0.13, which may be too low for meaningful exploration.  
Given the moderate gap and the need to maintain minimal exploration, I decide not to reinforce any option today.}
\end{agent}

The third response illustrates an indifference-band-like reasoning pattern. Although Route 2 has the lowest cost on the current day, the difference from Route 1 is only 0.1, which the agent treats as practically negligible. Instead of reinforcing Route 2 based on such a small difference, the agent chooses no reinforcement to avoid reacting to noise. This behavior is consistent with boundedly rational route-choice logic, where travelers may ignore cost differences that fall within an acceptable tolerance range.

\begin{agent}[$\omega_m^{t, \text{a-positive}}$ (the third example in the classic traffic assignment scenario)]
\textit{My current strategy is [0.895, 0.105] for routes 1 and 2.  
Today’s travel costs were [34.7, 34.6]. Route 2 had a slightly lower cost than Route 1 (difference of 0.1). From recent days: Route 1 has generally performed well, but today it was marginally worse than Route 2. Route 2 has had only one strong day (Day 2) and otherwise similar or higher costs, but today it was slightly better. The cost difference is extremely small (0.1), which is negligible and likely within normal daily variation.  
Given the minimal difference and to avoid reinforcing based on noise, I will not reinforce any option today.}
\end{agent}

\subsection{Highway tolling}
\label{app:tolling-response}

Next, we attach two pieces of the agent's reasoning under the highway tolling scenario, which help explain why a decoy-like pattern emerges in the LLM-guided dynamics. The key feature is that the agent does not evaluate each route only through an isolated scalar generalized cost. Instead, it often first compares the two tolled routes and recognizes that Route 3 is dominated by Route 2: Route 3 is both slower and more expensive. This semantic comparison makes Route 2 appear clearly attractive among the tolled alternatives and leads it to be repeatedly reinforced, while Route 1 is mainly evaluated as the slower but toll-free fallback. The following excerpts illustrate this reasoning pattern.

\begin{agent}[$\omega_m^{t, \text{a-positive}}$ (the first example in the highway tolling scenario)]
\textit{Current strategy (from yesterday's update). Route 1: 26.7\%; Route 2: 56.7\%; Route 3: 16.7\%. Today's travel times and costs.  Route 1: 40.1 min, \$0 toll; Route 2: 31.1 min, \$30 toll; Route 3: 37.0 min, \$34 toll.}

\smallskip
\textit{Recent experiences (last 2 days):}
\begin{itemize}[leftmargin=*, topsep=0pt, itemsep=0pt, parsep=0pt]
    \item[-] \textit{Day 1: Route 2 was 30.2 min, Route 3 was 37.1 min, Route 1 was 40.4 min.  }
    \item[-] \textit{Day 2 (today): Route 2 is 31.1 min, Route 3 is 37.0 min, Route 1 is 40.1 min.  }
\end{itemize}

\smallskip
\textit{Route 2 consistently fastest (30–31 min), Route 3 consistently middle (37 min), Route 1 slowest (40–41 min).   Route 3 costs more than Route 2 but is slower --- \textbf{clearly dominated by Route 2}. Route 1 is the slowest but free --- maybe worth keeping some weight for cost-saving.  Route 2 is reliable and worth the toll. Route 3 is not competitive. Route 1 is slow but free --- maybe I should try it again in case traffic improves, but so far it’s consistently slow. Given bounded rationality, I’ll stick with reinforcing Route 2 because it’s proven best so far. No need to increase Route 1 or 3 today.}  
\end{agent}

\begin{agent}[$\omega_m^{t, \text{a-positive}}$ (the second example in the highway tolling scenario)]
\textit{Current strategy:
Route 1: 27.9\%; Route 2: 70.4\%; Route 3: 1.8\%.
Today’s travel times and costs: Route 1: 40.1 min, \$0 toll; Route 2: 32.7 min, \$30 toll; Route 3: 37.0 min, \$34 toll.}

\smallskip
\textit{Recent experiences:}
\begin{itemize}[leftmargin=*, topsep=0pt, itemsep=0pt, parsep=0pt]
    \item[-] \textit{Route 2 has varied between about 30.2 and 33.1 minutes, and today dropped back to 32.7 minutes. } 
    \item[-] \textit{Route 3 has been consistently around 37.0 minutes.}
    \item[-] \textit{Route 1 has been consistently around 40 minutes. }
\end{itemize}

\smallskip
\textit{Route 2’s travel time dropped back to 32.7 minutes today — that’s nearly 7.4 minutes faster than Route 1, making the 30 HKD toll clearly worth it again. \textbf{Route 3 remains slower and more expensive than Route 2 — still dominated.} Route 1 is too slow when Route 2 is this fast — yesterday’s thought about cost-saving is less appealing when the time saving is this large. Given this variability, I’ll reinforce Route 2 today because it’s clearly the best option.  }
\end{agent}

\subsection{Transit transfer under denied-boarding risk}
\label{app:risk-response}

The following responses illustrate how the agent interprets denied-boarding risk. The key distinction lies in what the delayed outcome implies relative to the direct subway: in one setting, denied boarding makes the transfer option strictly worse, whereas in the other it erodes the transfer option’s advantage and signals declining reliability.

The first response shows reliability-sensitive reasoning in Setting A. The agent does not reject Option 1 simply because a bad outcome exists; rather, it reacts when the 52-minute outcome becomes frequent enough to make the transfer option feel unreliable.

\begin{agent}[$\omega_m^{t, \text{a-positive}}$ (the first example in the transit transfer under denied-boarding risk)]
\textit{My experiences this week:}
\begin{itemize}[leftmargin=*, topsep=0pt, itemsep=0pt, parsep=0pt]
    \item[-] \textit{Option 1: On 61\% of days, 30 minutes; on 39\% of days, 52 minutes.}  
    \item[-] \textit{Option 2: Always 45 minutes.}
\end{itemize}

\smallskip
\textit{\textbf{I'm starting to feel that Option 1's advantage is shrinking and the risk is growing.} It might be time to stop increasing Option 1. Option 2 is still slower, so I don't want to increase that either. I'll keep my strategy unchanged and not reinforce any option. } 
\end{agent}

The second response explains why Setting B has a much higher final share for Option 1. The agent recognizes that the bad transfer outcome no longer dominates the subway outcome.

\begin{agent}[$\omega_m^{t, \text{a-positive}}$ (the second example in the transit transfer under denied-boarding risk)]
\textit{This week's experiences:}
\begin{itemize}[leftmargin=*, topsep=0pt, itemsep=0pt, parsep=0pt]
    \item[-] \textit{Option 1: On 75\% of days, total time was 30 minutes. On 25\% of days, I was denied boarding and total time was 45 minutes.}  
    \item[-] \textit{Option 2: Always 45 minutes, no risk.}
\end{itemize}

\smallskip
\textit{\textbf{Even with the 25\% chance of a delay, Option 1 still seems better on average and often much faster (30 minutes vs 45). The 45-minute delay days match Option 2's time exactly, so Option 1 is never worse than Option 2 and often better.}}

\smallskip
Given that I've been using Option 1 more and it's still performing well, I'd like to increase its probability further to save more time on average.  
\end{agent}

However, Setting B does not collapse to an all-Option-1 outcome. Once the transfer option becomes heavily used, the agent starts to treat repeated denied boarding as a crowding and reliability signal, even though the delayed outcome is not worse than the subway.

\begin{agent}[$\omega_m^{t, \text{a-positive}}$ (the third example in the transit transfer under denied-boarding risk)]
\textit{This week, my strategy was heavily favoring Option 1. 
My recent experiences show that Option 1's success rate dropped: only 68\% of days were fast (30 min), and 32\% were delayed (45 min). That means nearly one in three days I ended up with the same travel time as Option 2, but with the hassle of waiting. Option 2 remains consistently 45 minutes with no risk.}

\smallskip
\textit{\textbf{Given that I increased Option 1 last week and now see a worsening of its reliability (more denied boardings), it seems that more people are also choosing it, causing congestion. Continuing to increase Option 1 might make the problem worse.} I feel it's better to shift back toward Option 2 to avoid the growing risk of delays. So I will increase the probability of Option 2 this week.  }
\end{agent}

\subsection{Multi-modal commuting}
\label{app:multi-modal-response}

Finally, we present several examples of the agent's reasoning in the multi-modal commuting scenario, which illustrates how different income classes interpret the same commuting options through different socioeconomic profiles. These examples help explain why the final mode-choice patterns differ across the low-, middle-, and high-income classes.

The first response illustrates how the middle-income representative agent interprets park-and-ride as a value-for-money compromise. The agent does not simply minimize monetary cost, which would favor transit, nor does it focus only on comfort, which would favor driving. Instead, it compares time, cost, and tolerable comfort together, and reinforces park-and-ride because it is close to driving in travel time while being noticeably cheaper. This helps explain why park-and-ride becomes the most attractive mode for the middle-income class in the main experiment.

\begin{agent}[$\omega_m^{t,\text{a-positive}}$ (the first example from the middle-income class)]
\textit{Today’s experiences:}
\begin{itemize}[leftmargin=*, topsep=0pt, itemsep=0pt, parsep=0pt]
    \item[-] \textit{Option 1 (transit): Time was about 58 minutes each way. The cost was low, but the experience remained poor: long exposed walk, two transfers, and a crowded second metro segment.}
    \item[-] \textit{Option 2 (driving): Time was about 35.7 minutes each way. The trip was comfortable, but the cost was high, and congestion made the travel time worse than usual.}
    \item[-] \textit{Option 3 (park-and-ride): Time was about 33 minutes each way. It was cheaper than driving and faster today, although the metro segment was somewhat crowded.}
\end{itemize}

\smallskip
\textit{Today, park-and-ride was clearly the better option: faster than driving, cheaper, and the metro crowding was manageable. Driving was slow due to congestion and expensive. Transit remains unappealing. Park-and-ride seems to be the most reliable and cost-effective choice overall.}

\textit{\textbf{I will increase park-and-ride's probability.}}

\end{agent}

The second and third responses show that the high-income agent does not choose park-and-ride mainly to save money. Instead, park-and-ride functions as a hedge against bad driving days. When highway congestion makes driving substantially slower, the agent reinforces park-and-ride. When driving time returns to normal and is close to park-and-ride, the agent favors driving because the comfort of a private car is worth the additional cost. This explains why both driving and park-and-ride retain substantial shares for the high-income class.

\begin{agent}[$\omega_m^{t,\text{a-positive}}$ (the second example from the high-income class: reinforcing park-and-ride)]
\textit{Today’s experiences:}
\begin{itemize}[leftmargin=*, topsep=0pt, itemsep=0pt, parsep=0pt]
    \item[-] \textit{Option 1 (transit): Around 58 minutes. It remained slow and uncomfortable, with walking and transfers.}
    \item[-] \textit{Option 2 (driving): 42.3 minutes. Highway congestion became much worse than in previous days.}
    \item[-] \textit{Option 3 (park-and-ride): Around 33 minutes. The metro segment was crowded, but the total travel time was much shorter than driving today.}
\end{itemize}

\smallskip
\textit{Driving had a bad day. Park-and-ride performed well today: fast travel time, lower cost, and a more tolerable metro segment. Given my high income, I value time and comfort. Today, park-and-ride clearly outperformed driving.}

\textit{\textbf{I want to increase the probability of Option 3 (park-and-ride).}}
\end{agent}

\begin{agent}[$\omega_m^{t,\text{a-positive}}$ (the third example from the high-income class: returning to driving)]
\textit{Today’s experiences:}
\begin{itemize}[leftmargin=*, topsep=0pt, itemsep=0pt, parsep=0pt]
    \item[-] \textit{Option 1 (transit): Still slow and uncomfortable.}
    \item[-] \textit{Option 2 (driving): About 35 minutes, back to normal travel time. It was comfortable and direct.}
    \item[-] \textit{Option 3 (park-and-ride): About 35 minutes, but the metro segment was crowded. It was cheaper than driving.}
\end{itemize}

\smallskip
\textit{Driving and park-and-ride were nearly identical in time today, but driving offers much more comfort. The cost difference is acceptable given my income. Park-and-ride is useful when driving is congested, but when travel times are similar, driving is more comfortable.}

\textit{\textbf{I want to increase the probability of Option 2 (driving).}}

\end{agent}

The last response illustrates the low-income agent's cost-sensitive reasoning. Transit remains attractive because of its low monetary cost, even though it is slow and uncomfortable. At the same time, the agent does not ignore comfort entirely. When the accumulated burden of walking, transfers, and crowding becomes salient, park-and-ride is reinforced as an occasional relief option rather than as a daily cost-minimizing choice. This explains why transit remains dominant for the low-income class, while park-and-ride still retains a substantial share.

\begin{agent}[$\omega_m^{t,\text{a-positive}}$ (the fourth example from the low-income class)]
\textit{Recent experiences:}
\begin{itemize}[leftmargin=*, topsep=0pt, itemsep=0pt, parsep=0pt]
    \item[-] \textit{Option 1 (transit): Very cheap, but still uncomfortable. The long exposed walk, transfers, and crowded metro are becoming tiring.}
    \item[-] \textit{Option 2 (driving): Much too expensive for my budget.}
    \item[-] \textit{Option 3 (park-and-ride): Still expensive, but less expensive than driving. It avoids the long walk and multiple transfers, and the drive portion is more comfortable.}
\end{itemize}

\smallskip
\textit{I have been favoring transit to save money, but the discomfort is accumulating. Park-and-ride is expensive, but it gives me a more comfortable and less tiring option. I cannot use it every day, but it may be worth using sometimes to get relief from the long walk and transfers.}

\textit{\textbf{I want to increase the probability of Option 3 (park-and-ride) slightly, while still keeping transit as the cost-saving fallback.}}
\end{agent}

These examples show that the same commuting mode can be interpreted differently across income classes. Park-and-ride is a value-for-money compromise for the middle-income class, a congestion hedge for the high-income class, and a comfort-relief option for the low-income class. These distinctions naturally arise from the representative agents' semantic interpretation of travel feedback under different socioeconomic profiles.

\section{Additional Robustness and Stability Checks}
\label{app:llm-configuration}

This appendix provides supplementary checks on two practical issues in the richer application settings: whether the baseline learning dynamics remain unstable as in the classic traffic assignment setting, and whether selected behavioral findings are robust to the choice of LLM core.

\subsection{Additional baseline results for highway tolling}
\label{app:baseline-tolling}

To examine whether the instability observed in the classic traffic assignment experiments persists under richer travel feedback, we test the fully LLM-driven baseline and the fully LLM-driven RL baseline in highway tolling Setting A without Road~3. Travelers choose between a toll-free road and a faster tolled road, using the scenario parameters in Section~\ref{sec:tolling} and the same LLM sampling settings. Figure~\ref{fig:baseline-tolling} reports one individual run for each baseline over 30 simulated days.

\begin{figure}[ht]
    \centering
    \includegraphics[width=0.7\textwidth]{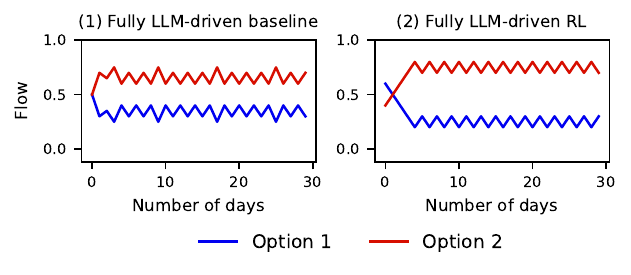}
    \caption{Selected individual trajectories of the fully LLM-driven and fully LLM-driven RL baselines under highway tolling Setting A without Road~3. Blue and red lines denote the shares of Options 1 and 2, respectively.}
    \label{fig:baseline-tolling}
\end{figure}

Both runs exhibit persistent oscillations and fail to settle into stable strategies over the simulated horizon, consistent with the classic traffic assignment results. These results further support explicit control over the numerical update.

\subsection{Robustness to the LLM core}
\label{app:llm-core}

We assess the robustness of the transit-risk findings in Section~\ref{sec:risk} by repeating both experiments using Qwen-3.8-Max in place of DeepSeek-V3.1. We use the same sampling configuration as in the main experiments, with temperature set to 0.2 and ``top\_p'' set to 1, and retain the same learning-rate schedule $\eta_m^t=c/(b+t)$ with $c=1$ and $b=3$. All other experimental settings remain unchanged.

We first repeat the four settings with successful/denied-boarding travel times of 42/52, 40/54, 38/56, and 36/58 minutes, respectively, while the subway travel time remains fixed at 45 minutes. Figure~\ref{fig:qwen-risk-results-additional} shows that the final transfer-option share rises from approximately 55\% in Setting A to 63\% in Setting B, before declining to about 55\% and 54\% in Settings C and D. This matches the non-monotonic pattern obtained with DeepSeek-V3.1 in Figure~\ref{fig:risk-additional}. Under both LLMs, the prospect of faster successful trips initially accompanies greater willingness to accept the transfer risk, while further widening of the travel-time gap reduces the transfer share. Thus, this non-monotonic behavioral response is not specific to DeepSeek-V3.1.

\begin{figure}[ht]
\centering
\includegraphics[width=\textwidth]{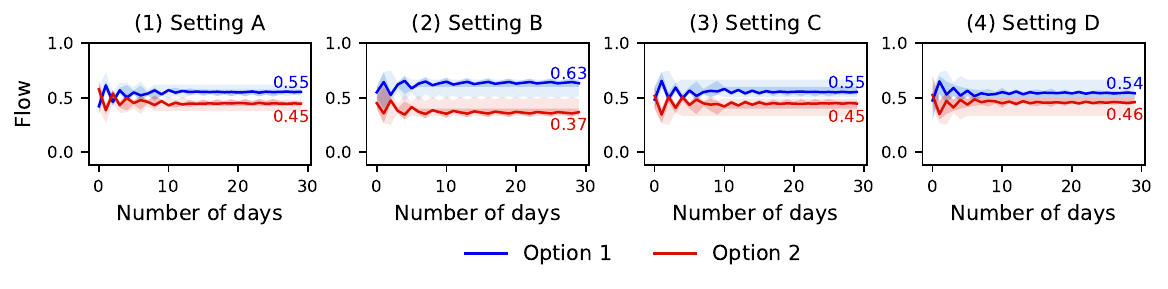}
\caption{Evolution of option shares under the LLM-guided RL dynamics using Qwen-3.8-Max for the four additional denied-boarding-risk settings with increasing separation between successful-boarding and denied-boarding travel times. Solid lines denote trimmed central-tendency trajectories; darker and lighter shaded areas denote the six-run and ten-run spreads, respectively.}
\label{fig:qwen-risk-results-additional}
\end{figure}

We next repeat the two targeted settings with a successful-boarding travel time of $\tau=30$ minutes and a headway of either $H=22$ or $H=15$ minutes. The denied-boarding trip therefore takes either 52 or 45 minutes, compared with the fixed 45-minute subway trip. This comparison tests whether the alternative LLM also distinguishes a delayed trip that is slower than the subway from one that takes the same time. As shown in Figure~\ref{fig:qwen-risk-results}, the final transfer-option share increases from approximately 52\% to 81\% when the delayed trip changes from taking 52 minutes to matching the 45-minute subway. The corresponding DeepSeek-V3.1 shares in Figure~\ref{fig:exp-risk-results} increase from about 56\% to 91\%. Both LLMs therefore assign a substantially larger share to the transfer option once denied boarding no longer makes it slower than the subway. This supports the robustness of the consequence-sensitive response, although the magnitude of the share increase differs between models.

\begin{figure}[ht]
\centering
\includegraphics[width=0.75\textwidth]{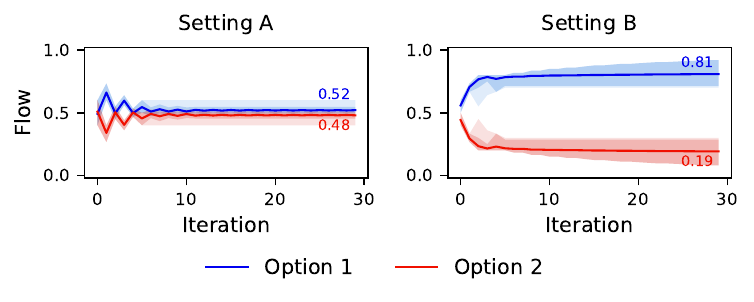}
\caption{Evolution of option shares under the LLM-guided RL dynamics using Qwen-3.8-Max for the two transit transfer settings. Solid lines denote trimmed central-tendency trajectories; darker and lighter shaded areas denote the six-run and ten-run spreads, respectively.}
\label{fig:qwen-risk-results}
\end{figure}

}

\vspace{10pt}

\section{Additional Ablation Checks}
\label{app:robustness}

\begin{figure}[ht]
    \centering
    \includegraphics[width=0.9\linewidth]{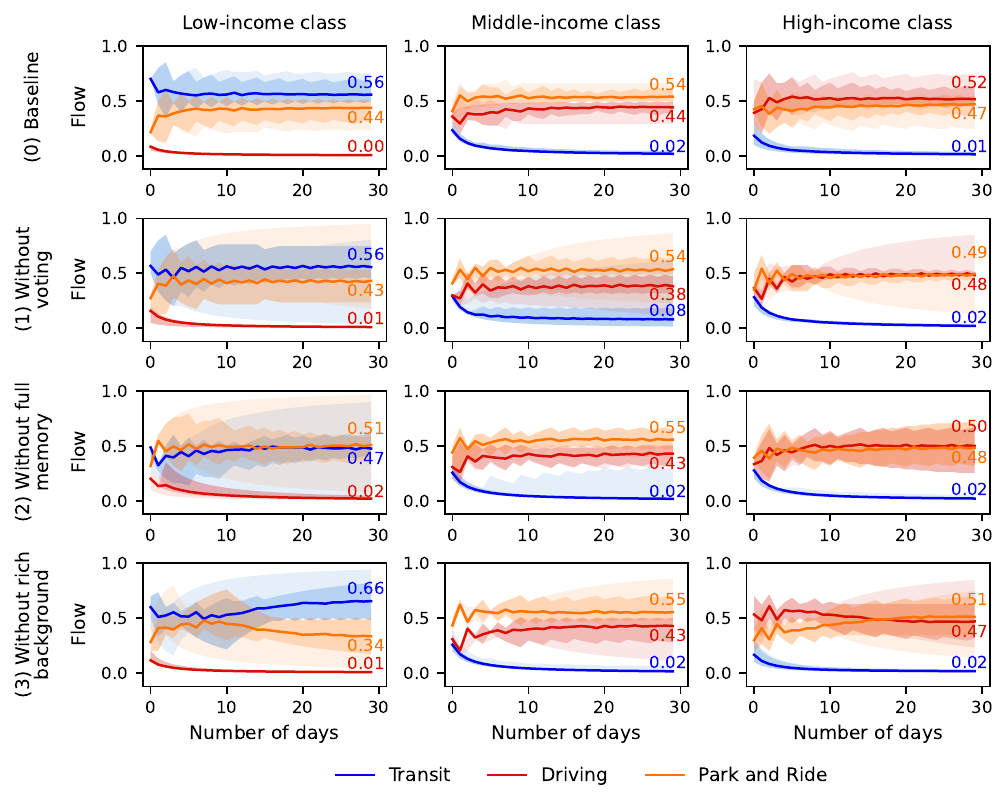}
    \caption{Ablation checks for the multi-modal commuting experiment. Row (0) reports the main specification. Row (1) removes the voting-based qualitative reasoning step. Row (2) uses bounded memory instead of the full dialog history. Row (3) removes rich socioeconomic background information and provides only income levels. Solid lines denote trimmed central-tendency trajectories; darker and lighter shaded areas denote the six-run and ten-run spreads, respectively.}
    \label{fig:exp3-ablation}
\end{figure}

This appendix reports three ablation checks for the multi-modal commuting experiment. The main specification uses the voting-based qualitative reasoning step, full dialog history, and rich socioeconomic background information. Here, we compare it with three variants: 
\begin{itemize}[topsep=0pt, itemsep=0pt, parsep=0pt]
    \item[-] removing the voting step and using a single LLM response;
    \item[-] using bounded memory instead of the full dialog history;
    \item[-] removing rich socioeconomic background information and providing only income levels. 
\end{itemize}
The results are shown in Figure~\ref{fig:exp3-ablation}. Row (0) reports the main specification as the baseline for comparison, while Rows (1)--(3) report the three ablation variants listed above. Within each row, the three columns correspond to the low-, middle-, and high-income classes, respectively. This layout allows us to compare how each implementation choice affects the mode-share dynamics of each traveler class.

\subsection{Removing the voting step}

When the voting step is removed, the broad qualitative pattern is largely preserved. The final shares become approximately $0.56/0.01/0.43$, $0.08/0.38/0.54$, and $0.02/0.48/0.49$ for the low-, middle-, and high-income classes, respectively. The main difference is that the trajectories become visibly more dispersed, especially for the middle- and high-income classes. This supports the interpretation that repeated LLM queries and majority voting mainly serve as a variance-control device. Without voting, an idiosyncratic single-day reinforcement decision may affect both the next-day mixed strategy and the subsequent dialog history, so such deviations can accumulate over time. Voting reduces this path-dependent amplification by making the daily reinforced set depend on the central tendency across multiple LLM responses rather than on one response alone. It does not fundamentally change the behavioral direction of the simulation, but reduces the influence of idiosyncratic single LLM responses on the daily reinforcement decision. 

This stability improvement, of course, comes at an additional computational cost:  the qualitative reasoning step requires 10 LLM queries for each representative agent on each simulated day. Therefore, voting should be viewed as an optional stability--cost trade-off. It is useful when a stable central tendency is desired for policy benchmarking, while the single-call version may be preferable when computational cost is more important.

\subsection{Using bounded memory}

When the full dialog history is replaced by bounded memory, the aggregate final shares remain broadly close to the baseline for the middle- and high-income classes. The low-income class changes more noticeably, with final shares shifting to approximately $0.47/0.02/0.51$.  A more important observation is that bounded memory produces more outlying runs, especially for the low-income class. A possible explanation is that the full dialog history helps keep the agent's decision logic and comparison baseline more consistent over time. 
The computational effect is mixed. As reported in Table~\ref{tab:token-runtime-summary} in Appendix~\ref{app:token}, bounded memory substantially reduces token usage in the multi-modal experiment: the average token usage per day decreases from 448.2K in the voting-based baseline to 197.2K under bounded memory. However, the runtime does not decrease correspondingly; in our implementation, the time per run is 3:01 for the baseline and 3:51 under bounded memory. This suggests that, in the current experimental setting, full dialog history is not the main runtime bottleneck. 
Therefore, bounded memory involves a trade-off. If a commercial API is used, reducing token usage can directly reduce monetary cost. However, if the main objective is to obtain stable and consistent repeated runs, bounded memory is not necessarily preferable, because it may increase outlying trajectories. For local LLM deployment, where monetary token cost is less direct and runtime may be dominated by model inference overhead rather than prompt length alone, the advantage of bounded memory may be even more limited. We therefore keep full dialog history as the main specification, while treating bounded memory as a useful option for very long-horizon simulations where token growth becomes a binding constraint.

\subsection{Removing rich background/profile information}

When the agents are given only income levels, without richer profile anchors such as car ownership, housing type, housing cost, and regular expenses, the most visible change occurs for the low-income class: transit becomes more dominant, with a final share of approximately $0.66$, while park-and-ride falls to about $0.34$. This change is intuitive. When only the label ``low income'' is provided, the agent has less contextual information for judging whether park-and-ride should be interpreted as an occasional comfort-relief option or simply as an expensive alternative to transit. More notably, the run-by-run variation becomes much more pronounced when rich background information is removed, especially for the low-income class. A possible explanation is that coarse income labels leave more room for the LLM to construct different implicit traveler profiles across repeated runs. Some runs interpret low income almost exclusively as cost minimization, while others allow more weight on fatigue and comfort. Richer socioeconomic anchors reduce this ambiguity and make the representative agent's reasoning more consistent across runs. 

This ablation has an important implication for future deployment. Representative LLM agents should not be specified only by coarse demographic labels such as income group. Instead, they should be provided with decision-relevant contextual information, ideally obtained from surveys, stated-preference data, observed behavior, or other empirical sources. Such information can improve not only behavioral interpretability, but also the stability and reproducibility of LLM-guided predictions.

\vspace{5pt}
\section{Token Usage and Computation Time}
\label{app:token}

This section reports token usage and runtime diagnostics for all experiments. For each experiment or setting, we report the number of representative classes $C$, the average number of alternatives per class Avg. $K$, the number of simulated days $T$, the number of LLM calls per class-day $q$, the average token usage per simulated day, and the recorded runtime per run. Note that the recorded runtime depends on API latency and server-side load. Nevertheless, experiments within the same scenario were run over approximately the same period, so the runtime comparisons within each scenario remain informative.

\begin{table}[ht]
\centering

\caption{Token usage and runtime diagnostics for all experiments}
\label{tab:token-runtime-summary}
\footnotesize
\begin{tabular}{lcccccc}
\toprule
Experiment / setting & $C$ & Avg. $K$ & $T$ & $q$ & Avg. tokens / day & Time / run \\
\midrule
\multicolumn{7}{l}{\textit{Classical traffic assignment}} \\
3N4L, Fully LLM-driven & 1 & 4.00 & 50 & 1 & 15.6K & 0:15:28 \\
3N4L, Fully LLM-driven RL & 1 & 4.00 & 50 & 2 & 22.9K & 0:28:45 \\
3N4L, LLM-guided RL (Rule 1) & 1 & 4.00 & 50 & 1 & 14.3K & 0:09:04 \\
3N4L, LLM-guided RL (Rule 2) & 1 & 4.00 & 50 & 1 & 13.7K & 0:09:58 \\
Hearn, Fully LLM-driven & 4 & 4.25 & 50 & 1 & 96.4K & 0:20:53 \\
Hearn, Fully LLM-driven RL & 4 & 4.25 & 50 & 2 & 110.8K & 0:30:47 \\
Hearn, LLM-guided RL (Rule 1) & 4 & 4.25 & 50 & 1 & 57.6K & 0:18:40 \\
Hearn, LLM-guided RL (Rule 2) & 4 & 4.25 & 50 & 1 & 66.1K & 0:22:31 \\
Sioux Falls, Fully LLM-driven & 528 & 2.10 & 50 & 1 & 1.81M & 1:16:13 \\
Sioux Falls, Fully LLM-driven RL & 528 & 2.10 & 50 & 2 & 2.57M & 1:24:50 \\
Sioux Falls, LLM-guided RL (Rule 1) & 528 & 2.10 & 50 & 1 & 3.39M & 1:22:15 \\
Sioux Falls, LLM-guided RL (Rule 2) & 528 & 2.10 & 50 & 1 & 3.12M & 2:22:15 \\
\midrule
\multicolumn{7}{l}{\textit{Highway tolling, LLM-guided RL (Rule 1)}} \\
Setting A (with decoy) & 1 & 3.00 & 30 & 1 & 7.9K & 0:07:26 \\
Setting A (without decoy) & 1 & 2.00 & 30 & 1 & 7.6K & 0:06:57 \\
Setting B (with decoy) & 1 & 3.00 & 30 & 1 & 8.2K & 0:08:07 \\
Setting B (without decoy) & 1 & 2.00 & 30 & 1 & 7.5K & 0:07:23 \\
Setting C (with decoy) & 1 & 3.00 & 30 & 1 & 7.8K & 0:05:32 \\
Setting C (without decoy) & 1 & 2.00 & 30 & 1 & 7.5K & 0:05:26 \\
\midrule
\multicolumn{7}{l}{\textit{Transit transfer under denied-boarding risk, LLM-guided RL (Rule 1)}} \\
Setting A & 1 & 2.00 & 30 & 1 & 10.2K & 0:01:17 \\
Setting B & 1 & 2.00 & 30 & 1 & 10.1K & 0:01:53 \\
\midrule
\multicolumn{7}{l}{\textit{Multi-modal commuting, LLM-guided RL (Rule 1)}} \\
Voting-based & 3 & 3.00 & 30 & 10 & 448.2K & 0:03:01 \\
\midrule
\multicolumn{7}{l}{\textit{Additional ablation checks, LLM-guided RL (Rule 1)}} \\
Without voting & 3 & 3.00 & 30 & 1 & 46.9K & 0:02:50 \\
Without full memory & 3 & 3.00 & 30 & 10 & 197.2K & 0:03:51 \\
Without rich background & 3 & 3.00 & 30 & 10 & 441.3K & 0:03:00 \\
\bottomrule
\end{tabular}
\end{table}

\end{document}